\documentclass[prd,showpacs,nofootinbib,preprint, 10 pt]{revtex4}
\usepackage{amsmath,graphicx,xcolor,epsfig,slashed}
\usepackage{appendix}
\usepackage{tabularx}
\usepackage{float}
\usepackage{pdfpages}
\usepackage{epsfig}
\usepackage{epstopdf}
\usepackage{subfigure}
\usepackage{caption}
\usepackage{subcaption}
\usepackage{multirow}
\usepackage{hyperref}

\begin{document}
\begin{center}
{\bf\Large\boldmath
Imprinting New Physics by using Angular profiles of the FCNC process
$B_{c}\to D_{s}^{*}\left(\to \;D_{s}\pi\right)\ell^{+}\ell^{-}$}\\[5mm]
\par\end{center}
\begin{center}
\setlength {\baselineskip}{0.2in}
{Hira Waseem, Abdul Hafeez }\\[5mm]
{\it Department of Physics, Quaid-i-Azam University, Islamabad 45320, Pakistan.}
\\[5mm]
\end{center}
{\bf Abstract}\\[5mm]
The decays governed by the flavor-changing-neutral-current transitions (FCNC), such as $b\to s\ell^{+}\ell^{-}$, provide an important tool to test the physics in and beyond the Standard Model (SM). This work focuses on investigating the FCNC process $B_{c}\to D_{s}^{*} \left(\to D_{s}\pi\right)\ell^{+}\ell^{-}(\ell=e,\mu,\tau)$. Being an exclusive process, the initial and final state meson matrix elements involve the form factors, which are non-perturbative quantities and need to be calculated using specific models. By using the form factors calculated in the covariant light-front quark model, we analyze the branching fractions and angular observables such as the forward-backward asymmetry $A_{FB}$, polarization fractions (Longitudinal and transverse) $F_{L(T)}$, CP asymmetry coefficients $A_{i}$ and CP-averaged angular coefficients $S_{i}$, both in the SM and some new physics (NP) scenarios. Some of these physical observables are a potential source of finding the physics beyond the SM and help us distinguish various NP scenarios.

\maketitle

\section{Introduction}
 Several results from the last few decades have some $(1-3)\sigma$ disagreement with the Standard Model results, and the FCNC processes involving $b \to s$ are the pertinent ones.  The Glashow–Iliopoulos–Maiani (GIM) mechanism allows these transitions at loop level in the Standard Model (SM), and due to their strong suppression within the SM, exclusive and inclusive $b\to s\ell^{+}\ell^{-}$ decays are the potential probes of short-distance physics. Also, their sparkling sensitivity to the new physics (NP) particles makes them harbinger to study the NP indirectly \cite{LHCb:2021vvq}. In 2013, the observation of tension by LHCb led to the assumption of the presence of New Physics \cite{LHCb:2021auc}. In 2014, another dilemma was confronted within the SM, namely the suppression of the ratio $R_{K}$ for $B \to K \ell^{+}\ell^{-} (\ell=e,\mu)$ at low Dilepton invariant mass and for the consistent description of these anomalies resulted into the presence of new physics \cite{LHCb:2017iph}. Additionally, the calculated value of the branching ratio of $B_{(s)} \to \phi \mu^{(+)} \mu^{(-)}$ \cite{Bicudo:2021qxj} was small compared to the SM observations \cite{Ali:2019roi,  LHCb:2015wdu}. 
 
 One of the most optimised observable $\left(\mathcal{P}_5\right)$ in $B\to K^*\mu^{+}\mu^{-}$ decay has shown the mismatch with the SM predictions \cite{Descotes-Genon:2012isb, Descotes-Genon:2013vna}. The Yukawa sector of the SM was also scrutinized by measuring the Lepton Flavor Universality (LFU) ratio $\mathcal{R}_{K^{(*)}}\equiv \frac{\mathcal{B}\left(B\to K^{(*)}\mu^{+}\mu^{-}\right)}{\mathcal{B}\left(B\to K^{(*)}e^{+}e^{-}\right)}$ in various bins of the transferred square momentum, i.e. $q^2=\left(p_{\ell^+}+p_{\ell^-}\right)^2$ and almost $3\sigma$ deviations from the SM predictions were recorded at the LHCb \cite{LHCb:2017avl, LHCb:2019hip, LHCb:2021trn}. However, the situation changed after the latest measurements of LHCb in low and central $q^2$ region \cite{LHCb:2022qnv, LHCb:2022vje}, making the LFU ratio in agreement with the SM predictions. Motivated by these tensions, several theoretical studies were performed for the complimentary exclusive decays $B\to \left(K_{1}\left(1270,1430\right), K^{*}_{2} \left(1430\right), f^{\prime}_{2} \left(1525\right)\right)\ell^{+}\ell^{-}$ in the SM and various NP models
 \cite{Ishaq:2013toa, Munir:2015gsp, Huang:2018rys, Sikandar:2019qyb, MunirBhutta:2020ber, Das:2018orb, Mohapatra:2021izl, Rajeev:2020aut, Sikandar:2022iqc}.
The semi-leptonic decays involving tauons have received less attention than the muon and electron in the final state. This situation has now changed, and after the experimental improvements, these decays are in the limelight now \cite{LHCb:2017myy, Vobbilisetti:2023fnz, Celani:2021hni, Belle:2021ecr, Li:2020bvr}. Theoretically, to scrutinize the various NP models, the semileptonic B-meson decays occurring through the FCNC transition $b\to (s,d)\tau^{+}\tau^{-}$ have been investigated in several studies, see e.g., \cite{Alok:2024cyq, Becirevic:2024vwy, Alok:2023yzg, Ali:2023kvz, Ho:2022ipo, Maji:2021tbl, Cornella:2021sby, Cornella:2020aoq, Huber:2020vup}. 
In contrast to the ordinary $B$ meson decays, the $B_c$ decay may proceed through the weak decay of either of its heavy constituents, i.e., $b$ or $c$, and the other will play the role of spectator. It can also decay virtually to $W^+$ boson, and its lifetime is almost one-third of that of the $B^0,\; B^+$ \cite{LHCb:2014ilr, LHCb:2014mvo, LHCb:2014glo}. 
Recently, using the $9 fb^{-1}$ data if the proton-proton collision at the LHC, the LHCb has performed the searches for $B_{c}^{+}\to D_{s}^{+} \mu^{+} \mu^{-}$ decay, but have not observed any significant signal in the non-resonant dimuonic mode. The upper limits on $\frac{f^{c}}{f^{u}}\times\mathcal{B}\left(B_{c}^{+}\to D_{s}^{+}\mu^{+}\mu^{-}\right)< 9.6\times 10^{-8}$ is set with $95\%$ c.f. Here, $f^{c}$ and $f^{u}$ are the fragmentation fractions of a $B$ meson having $c$ and $u$ quarks \cite{LHCb:2023lyb} . As $D_{s}^{*}$ needs to be rebuilt from the $D_{s}$ meson experimentally, it will be a little tough to measure $B_{c}\to D_{s}^{*}\ell^{+}\ell^{-}$ but the situation can be made better with upgrade in high luminosity of LHCb in future.

On theoretical fronts, the semileptonic $B_{c}\to D_{s}^*$ decay was studied in several approaches, e.g., light-front quark model (LFQM) \cite{Geng:2001vy}, the perturbative QCD (pQCD) approach \cite{Wang:2014yia}, the QCD sum rule \cite{Kiselev:2002vz, Azizi:2008vv}, the constituent quark model (CQM) \cite{Geng:2001vy}, etc.  In the SM, the branching ratio for the electron and muon mode is calculated to be of the order of $10^{-8}$ while for the tau is $10^{-9}$, particularly, various physical observables of $B_c\to D_{s}^*\mu^{+}\mu^{-}$ decay have been calculated in the SM and beyond in Refs. \cite{Dutta:2019wxo,Mohapatra:2021ynn,Zaki:2023mcw}. Including the $\tau$ leptons in the final state, $B_{c} \to D_{s}^{*}\ell^{+}\ell^{-}$ has been studied in \cite{Li:2023mrj}, where Li \textit{et al.} have calculated the form factors in the covariant LFQM approach. Also, the full calculation of the angular distribution of the quasi-four-fold distribution $B_{c} \to D_{s}^{*}\left(D\pi\right)\ell^{+}\ell^{-}$ have been done to study various physical observables.

As a follow-up of the study presented in \cite{Li:2023mrj}, using their form factors, we will first analyze the branching fractions and angular observables such as the forward-backwards asymmetry $A_{FB}$, polarization fractions $F_{L(T)}$, CP averaged angular observables ($S_{i}$, $A_{i}$) both in the SM and beyond. For the physics beyond the SM, we used the latest model-independent global fit to $b\to s \ell^{+}\ell^{-}$ observables which include the latest measurements LHCb observables of $B_s \to \mu^{+}\mu^{-}$, $B_s\to \phi \mu^{+}\mu^{-}, \; R_{K, K_S}$ and $R_{K^*}$ and added the 254 observables to find the pattern of the NP that successfully explain the data \cite{Alguero:2021anc}. We hope our findings will complement the various asymmetries observed in FCNC decays.

The structure of this paper is organized as follows: Following the introductory section, we present the angular distributions of the quasi-four body decays $B_{c}\; \to \; D_{s}^{*} \left(\to D_{s} \pi\right) \ell^{+} \ell^{-}$ in section \ref{sec:2} and \ref{sec:3}. In section \ref{sec:4}, we introduce the form factors derived by the covariant light-front quark model in \cite{Li:2023mrj}. Section \ref{sec:5} presents numerical results for several observables in the Standard Model and next-generation physics using NP models. Finally, this letter ends with an outline.

\section{Effective Hamiltonian in SM and BSM}\label{sec:2}
 The low energy effective Hamiltonian for $b \to s \ell^{+} \ell^{-}$ in the model-independent way is written as \cite{Isgur:1990yhj}, 
 \begin{equation}
\mathcal{H}_{eff} = -\frac{4G_F}{\sqrt{2}}\left[\lambda_{u}\left(C_1\left (\mathcal{O}_1^{u}-\mathcal{O}_{1}^{c}\right)+C_2\left(\mathcal{O}_{2}^{u}-\mathcal{O}_{2}^{c}\right)\right)+\lambda_{t}\sum_{i \in J} C_{i}\mathcal{O}_{i}\right],\label{eq:01}
 \end{equation}
where $\lambda_{q}$ denotes $V_{qb}V_{qs^{*}}$ and $J=\left(1_{c}, 2_{c}, 3...8,7^{\left(\prime\right)},9l^{\left(\prime\right)},10l^{\left(\prime\right)},\;Sl^{\left(\prime\right)},Pl^{\left(\prime\right)},Tl^{\left(\prime\right)}\right)$. $V_{ij}$ stands for the CKM matrix elements and the Fermi constant is represented by $G_F=1.16637\times 10^{-5}$ GeV$^{-2}$, $C_{i}(\mu)$ are the Wilson coefficients that tell about the strength of the interaction and $\mathcal{O}_{i}$ are the four fermion operators while $\mu$ is the renormalization scale. Particularly, $\mathcal{O}_{1,2}$ are the current-current operators, $\mathcal{O}_{3-6}$ are the penguin operators, the $\mathcal{O}_{7,8}$ are the electromagnetic operators and $\mathcal{O}_{9,10}$ are the semi-leptonic operators and their corresponding Wilson coefficients describe the coupling strength between the respective quarks and charged leptons at the factorization scale $\mu=m_b$. Short-distance physics is encoded in Wilson coefficients of higher dimension operators. In SM, the effective Hamiltonian contains 10 operators with specific chiralities due to the V-A structure of weak interactions where the heavy degrees of freedom are integrated out, and we are left with only the operators set, describing the long-distance physics \cite{Coleangelo}.

Here, the effective Hamiltonian used has the following form,
\begin{eqnarray}
\mathcal{H}_{eff} \left(b \to s \ell^{+} \ell^{-}\right ) &=& -\frac{4G_F}{\sqrt{2}} V_{tb}V_{ts}^{*} \frac{\alpha_{e}}{4 \pi}\Bigg[\bar{s}\left(C_{9}^{eff}(q^{2},\mu)\gamma^{\mu} P_{L}-\frac{2m_{b}}{q^{2}} C_{7}^{eff}(\mu)i\sigma^{\mu \nu} q_{\nu}P_{R}\right)b(\bar{\ell}\gamma_{\mu}\ell)\notag\\
&&+C_{10}(\mu)\left(\bar{s}\gamma^{\mu} P_{L}b\right)\left(\bar{\ell}\gamma^{\mu}\gamma_{5}\ell\right)\Bigg]  \label{eq:02}
\end{eqnarray}
where $P_{L}$ and $P_{R}$ stands for the left and right projection operators, $\sigma^{\mu \nu}=i(\gamma^{\mu}\gamma^{\nu}-\gamma^{\nu}\gamma^{\mu})/2$ and $\alpha_{e}$ stands for the electromagnetic coupling which is $\frac{1}{137}$. 
The operators have the form, 
\begin{eqnarray}
    \mathcal{O}_{7}&=&\frac{e}{16\pi^{2}}m_{b}\left(\bar{s}\sigma_{\mu\nu}P_{R}b\right)F^{\mu\nu}, \notag\\
    \mathcal{O}_{9}&=&\frac{e^2}{16\pi^{2}}\left(\bar{s}\gamma_{\mu}P_{L}b\right)\left(\bar{l}\gamma^{\mu}l\right), \notag\\
    \mathcal{O}_{10}&=&\frac{e^2}{16\pi^{2}}\left(\bar{s}\gamma_{\mu}\gamma_{5}P_{L}b\right)\left(\bar{l}\gamma^{\mu}l\right). 
\end{eqnarray}
Here, operator $\mathcal{O}_{7}$ represents the interaction between photons and quarks while $\mathcal{O}_{9,10}$ represents the interaction between quarks and leptons, respectively. $C_{7}^{eff}$ and $C_{9}^{eff}$ are represented by 
\begin{align}
C_{7}^{eff}(\mu)& = C_{7}(\mu)+C^{\prime}_{b \to s \gamma}(\mu) \notag\\
C_{9}^{eff}(q^2,\mu) &= C_{9}(\mu)+C_{9,pert}(q^2,\mu)+C_{9,c\bar{c}}(q^2,\mu), \label{eq:3}
\end{align}
where $C^{\prime}_{b \to s \gamma}(\mu)$ results from the following interaction $b \to s c \bar{c} \to s\gamma$ \cite{Falk:1990yz}. $C_{9,pert}(q^2,\mu)$ and $C_{9,c\bar{c}}(q^2,\mu)$ represent the short and long-distance contributions respectively as $C_{9,pert}(q^2,\mu)$ results from the one-loop matrix element of the four fermi quark operators $\mathcal{O}_{1-6}$. It can be calculated at leading order within perturbation theory while $C_{9,c\bar{c}}(q^2,\mu)$ relies upon the external hadron states. Putting everything together \cite{M.J.Aslam}
\begin{equation}
C^{\prime}_{b \to s \gamma}\left(\mu\right)=i\alpha_{s} \left[\frac{2}{9}\eta^{\frac{14}{23}}\left( \frac{y_{t}\left(  y_{t}^2-5y_{t}-2 \right )}{8\left(y_{t}-1\right)^3}+\frac{3y_{t}^{2}\log{y_{t}}}{4\left(y_{t}-1\right)^{4}}-0.1687 \right)-0.03\times C_{2}(\mu)\right], \label{eq:4}
\end{equation}
with $y_{t}=\frac{m_{t}^2}{m_{W}^2}$, $\eta=\frac{\alpha_{s}(m_{W})}{\alpha_{s}(\mu)}$ and the mass scale $\mu=m_{b}$.
\begin{eqnarray}
C_{9,pert}(\hat{s},\mu)&=&0.124\omega \left(\hat{s}\right)+g\left(\hat{m_{c}},\hat{s}\right)C_{\mu}+\lambda_{\mu}\Big[ g\left(\hat{m_{c}},\hat{s}\right)-g\left(0,\hat{s}\right)\Big]\left(3C_{1}\left(\mu\right)+C_{2}\left(\mu\right)\right)-\frac{1}{2}g\left(0,\hat{s}\right)\left(C_{3}\left(\mu\right)+3C_{4}\left(\mu \right)\right)-\notag\\
&&\frac{1}{2}g\left(1,\hat{s}\right)\left(4C_{3}(\mu)+4C_{4}(\mu)+3C_{5}(\mu)+C_{6}(\mu)\right)+\frac{2}{3}C_{3}(\mu)+\frac{2}{9}C_{4}(\mu)+\frac{2}{3}C_{5}(\mu)+\frac{2}{9}C_{6}(\mu),
\end{eqnarray}
where $\hat{s}=\frac{q^2}{m_{b}^2}$ and $\hat{m}_{c}=\frac{m_{c}}{m_{b}}$ and $C\left(\mu\right)=3C_{1,3,5}\left(\mu \right)+C_{2,4,6}\left(\mu\right)$. The Wilson coefficients are chosen at $\mu=m_{b}$ and their numerical values read as $C_{1}= -0.226, C_{2}=1.096, C_{3}=0.01, C_{4}=-0.024, C_{5}=0.007, C_{6}=-0.028, C_{7}=-0.305, C_{8}=-0.15, C_{9}=4.186, C_{10}=-4.559,$ \cite{Virto}.
In the Wolfenstein representation, the $\lambda_{\mu}$ has the following form \cite{Ligeti:1993hw}
\begin{equation}
\lambda_{\mu}\approx -\lambda^2 \left(\rho-i\eta \right), \label{eq:6}
\end{equation}
and $\lambda=0.22500\pm0.00067$. Here the function $\omega\left(\hat{s}\right)$ is defined as \cite{Luke:1990eg}:
\begin{eqnarray}
\omega\left(\hat{s}\right)&=&-\frac{2}{9}\pi^{2}+\frac{4}{3}\int_{0}^{\hat{s}}\frac{\log \left (1-u \right)}{u}du-\frac{2}{3}\log{\hat{s}}\log{\left(1-\hat{s}\right)}-\frac{5+4\hat{s}}{3\left(1+2\hat{s}\right)}\log\left(1-\hat{s}\right)-\frac{2\hat{s}\left(1+\hat{s}\right)\left(1-2\hat{s}\right)}{3\left(1-\hat{s}\right)^{2}\left(1+2\hat{s}\right)}\log \hat{s}+\notag\\
&&\frac{5+9\hat{s}-6\hat{s}^{2}}{6\left(1-\hat{s}\right)\left(1+2\hat{s}\right)}, \label{eq:7}
\end{eqnarray}
and the functions $g(z,\hat{s})$ have the form \cite{Neubert:1991td,Shifman:1994jh}
\begin{eqnarray}
g\left(z,\hat{s}\right)&=-\frac{8}{9}\log\left(z\right)+\frac{8}{27}+\frac{4}{9}y-\frac{2}{9}\left(2+y\right)\sqrt{|1-y|}\times \log\left|\frac{1+\sqrt{1-y}}{1+\sqrt{1-y}|-i\pi}\right| \hspace{0.5 cm}\text{for} \hspace{0.5 cm}y\equiv\frac{4z^{2}}{\hat{s}} <1\notag\\
g\left(z,\hat{s}\right)&=-\frac{8}{9}\log(z)+\frac{8}{27}+\frac{4}{9}y-\frac{2}{9}\left(2+y\right)\sqrt{|1-y|}\times 2\arctan\frac{1}{\sqrt{y-1}} \hspace{0.5 cm} \text{for} \hspace{0.5 cm} y \equiv \frac{4z^{2}}{\hat{s}}>1, \label{eq:8}
\end{eqnarray}
which for $q^2=0$ becomes:
\begin{table}[h]
\begin{tabular}{|l|l|l|l|}
\hline\hline
$V_{i}$ & $m_{V_{i}}$ & $\Gamma_{V_{i}}$ & $\mathcal{B}(V_{i}\to \ell^{+}\ell^{-})\times 10^{-5}$ \\
\hline
$\rho$ & 0.775 & 149 & $4.635$ \\
$\omega$ & 0.783 & 8.68 &  $7.380$ \\
$\phi$ & 1.019 & 4.249 & $2.915\times 10^{+1}$ \\
$J/\psi$ & 3.097 & 0.093 & $5.966 \times 10^{+3}$ \\
$\psi(2S)$ & 3.686 & 0.294 & $7.965 \times 10 ^{+2}$ \\
$\psi_{(3770)}$ & 3.774 & 27.2 & $9.6\times 10^{-1}$ \\
$\psi_{(4040)}$ & 4.039 & 80 & $1.07 \times 10^{+1}$ \\
$\psi_{(4160)}$ & 4.191 & 70 & $6.9 \times 10^{-1}$ \\
\hline\hline
\end{tabular}
\caption{Properties of resonances and the values of input parameters involved in effective Wilson coefficients \cite{Li:2023mrj}.}
\label{table 1}
\end{table}
\begin{equation}
g(0,\hat{s})=\frac{8}{27}-\frac{8}{9}\log \left( \frac{m_{b}}{\mu}\right)-\frac{4}{9}\log\hat{s}+\frac{4}{9}i\pi . \label{eq:9}
\end{equation}
The $C_{9,c\bar{c}\left(q^2,\mu\right)}$ has the contributions from the intermediate light vector mesons and vector charmonium states\cite{Czarnecki:1996gu}
\begin{eqnarray}
C_{9,c\bar{c}\left(q^2,\mu\right)}&=&-\frac{3\pi}{\alpha_{e}^2}\Bigg [C\left(\mu\right)\sum_{V_{i}=J/\psi,\psi\left(2S\right),\dots}\frac{m_{V_{i}}\mathcal{B}\left(V_{i}\to l^{+}l^{-}\right)\Gamma_{V_{i}}}{q^2-m_{V_{i}}^2+i m_{V_{i}}\Gamma_{V_{i}}}-\lambda_{u}g(0,\hat{s})\Big(3C_{1}(\mu)+C_{2}(\mu)\Big)\notag\\
&&\sum_{V_{j}=\rho,\omega,\dots}\frac{m_{V_{j}}\mathcal{B}(V_{j}\to l^{+}l^{-})\Gamma_{v_{j}}}{q^2-m_{V_{j}}^2+im_{V_{j}}\Gamma_{V_{j}}} \Bigg] .
\label{eq:10}
\end{eqnarray}
Where $m_{V_{i}}$ and $\Gamma_{V_{i}}$ represent the mass in GeV and the total decay rate of the particular resonance particle in MeV. These values are taken from the Particle Data Group \cite{PDG2020} and represented in Table \ref{table 1}.

\section{Angular distributions and the observables}\label{sec:3}
\begin{figure}[h]
    \centering
    \includegraphics[width=4.5in,height=4.5in]{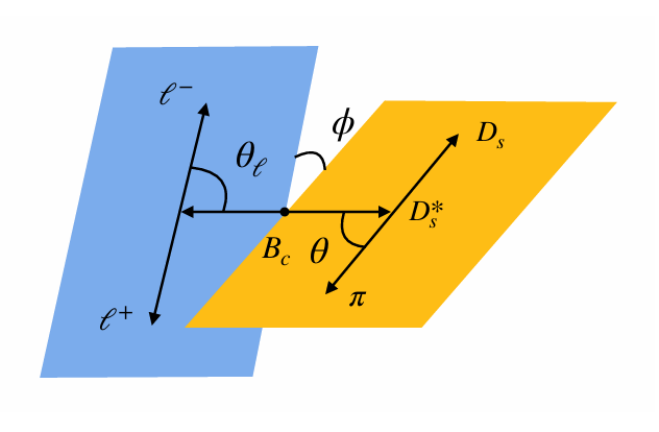}
    \caption{Scattering Kinematics of quasi-four fold distribution \cite{Li:2023mrj}.}
    \label{fig: Kinematics of quasi four body decay}
\end{figure}

By definition of the transition probability amplitude
\begin{eqnarray}
\langle D_{s}^{*}(p^{\left(2\right)})|\bar{s}\gamma_{\mu}b|B_{c}(p^{(1)})\rangle&=&\epsilon_{\mu \nu \alpha \beta} \epsilon^{*\nu}P^{\alpha}q^{\beta}g(q^2),\notag\\
\langle D_{s}^{*}(p^{(2)})|\bar{s}\gamma_{\mu}\gamma_{5}b|B_{c}(p^{(1)})\rangle&=&-i\left[\epsilon^{*}_{\mu}f\left(q^2\right)+\epsilon^{*}.P\left(P_{\mu}a_{+}\left(q^2\right)+q_{\mu}a_{-}\left(q^2\right)\right)\right], \label{eq:11}
\end{eqnarray}
where $P_{\mu}=p_{\mu}^{(1)}+p_{\mu}^{(2)}$ and $q_{\mu}=p_{\mu}^{(1)}-p_{\mu}^{(2)}$. Also, $\epsilon$ represents the polarisation vector of the $D_{s}^{*}$ meson. The expression of amplitudes in the Bauer-Stech-Wirbel (BSW) form \cite{Mannel:2023pdg} as follows:
\begin{eqnarray}
\langle D_{s}^{*}(p^{(2)})|\bar{s}\gamma_{\mu}b|B_{c}(p^{(1)})\rangle &=&-\frac{1}{m_{1}+m_{2}}\epsilon_{\mu \nu \alpha \beta} \epsilon^{*\nu}P^{\alpha}q^{\beta}V(q^2),\notag\\
\langle D_{s}^{*}(p^{(2)})|\bar{s}\gamma_{\mu}\gamma_{5}b|B_{c}(p^{(1)})\rangle&=&i\left[\left(m_{1}+m_{2}\right)\epsilon^{*}_{\mu}A_{1}\left(q^2\right)-\frac{\epsilon^{*}.P}{m_{1}+m_{2}} P_{\mu}A_{2}\left(q^2\right)-2m_{2}\frac{\epsilon^{*}.P}{q^2}q_{\mu}\left[A_{3}\left(q^2\right)-A_{0}\left(q^2\right)\right]\right], \label{eq:12}
\end{eqnarray}
where the mass of the $B_{c}$ meson is $m_{1}$, while the mass of the $D_{s}^{*}$ meson is $m_{2}$. Additionally, tensor current amplitude is defined as \cite{Kiselev:2001fw}
\begin{eqnarray}
\langle D_{s}^{*}(p^{(2)})|\bar{s}i\sigma^{\mu \nu}q_{\nu}b|B_{c}(p^{(1)})\rangle &=& T_{1}(q^2)\epsilon_{\alpha \beta \mu \nu}\epsilon^{*}_{\nu}P^{\alpha}q^{\beta}, \notag\\
\langle D_{s}^{*}(p^{(2)})|\bar{s}i\sigma^{\mu \nu}q_{\nu}\gamma{5}b|B_{c}(p^{(1)})\rangle &=&iT_{2}\left(q^2\right)\Big[\left(m_{1}^2-m_{2}^2\right)\epsilon^{*}_{\mu}-\epsilon^{*}.qP_{\mu}\Big]+iT_{3}\left(q^2\right)\epsilon^{*}.q\left[q_{\mu}-\frac{q^2 P_{\mu}}{m_{1}^2-m_{2}^2}\right]. \label{eq:13}
\end{eqnarray}
Here, $V(q^2), A_{1}\left(q^2\right), A_{2}\left(q^2\right), A_{3}\left(q^2\right), A_{0}\left(q^2\right), T_{1}\left(q^2\right), T_{2}\left(q^2\right), T_{3}\left(q^2\right)$ are the form factors \cite{Manohar:1992nd}, \cite{Caprini:1997mu},\cite{ParticleDataGroup:2022pth}. By using the above definitions of the matrix elements, the invariant amplitude is written in the following form,
\begin{eqnarray}
\mathcal{M}(s_{l^{+}},s_{l^{-}})&=& \langle D_{s}\pi ;l^{+}(s_{l^{+}})l^{-}(s_{l^{-}})|\mathcal{H}_{eff}|B_{c}\rangle\notag\\
&=&\sum_{s_{v}}\frac{I}{k^{2}-m_{D_{s}^{*}}^2}\mathcal{M}_{D_{s}^{*}\to D_{s} \pi }\left(s_{V}\right)\langle D_{s}^{*}\left(s_{V}\right)l^{+}\left(s_{l^{+}}\right)l^{-}\left(s_{l^{-}}\right)|\mathcal{H}_{eff}|B_{c}\rangle, \label{eq:14}
\end{eqnarray}
\text{where} $\mathcal{M}_{D_{s}^{*}\; \to \;D_{s}\pi}$ \text{is}
\begin{eqnarray}
\mathcal{M}_{D_{s}^{*}\; \to\; D_{s}\pi}=-ig_{D_{s}^{*} D_{s}} \epsilon^{\mu}p^{\pi}_{\mu}. \label{eq:15}
\end{eqnarray}
Using the effective Lagrangian approach for the calculation, the amplitude has the following form,
\begin{eqnarray}
\mathcal{M}\left(s_{l^{+}},s_{l^{-}}\right)&=&\sum_{s_{v}}\frac{I}{k^{2}-m_{D_{s}^{*}}^2}\mathcal{M}_{D_{s}^{*}\;\to \;D_{s} \pi }\left(s_{V}\right)\Bigg[C_{9}^{eff}H^{V-A}\left(s_{V},t\right)L^{V}\left(s_{l^{+}},s_{l^{-}},t\right)-\frac{2m_{b}}{q^2}C_{7}^{eff}H^{T+T5}\left(s_{V},t\right)\notag\\
&&L^{V}\left(s_{l^{+}},s_{l^{-}},t\right)+C_{10}H^{V-A}\left(s_{V},t\right)L^{A}\left(s_{l^{+}},s_{l^{-}},t\right)-\sum_{\lambda=0,\;\pm}\Bigg(C_{9}^{eff}H^{V-A}\left(s_{V},\lambda\right)L^{V}\left(s_{l^{+}},s_{l^{-}},\lambda \right)-\notag\\
&&\frac{2m_{b}}{q^2}C_{7}^{eff}H^{T+T5}\left(s_{V},\lambda \right)L^{V}\left(s_{l^{+}},s_{l^{-}},\lambda \right)+C_{10}H^{V-A}\left(s_{V},\lambda \right)L^{A}\left(s_{l^{+}},s_{l^{-}},\lambda \right)\Bigg)\Bigg].  \label{eq:16}
\end{eqnarray}
In this equation, $N=\frac{4G_{F}\alpha_{e}V_{tb}V_{ts}^{*}}{ 4\pi \sqrt{2}}$ and a factor of $\frac{1}{2}$ comes from the right and left-hand projection operators mentioned in Eq.(\ref{eq:02}). Finally, by using effective Hamiltonian as mentioned in Eq. (\ref{eq:02}), the simplified form  of angular distribution as deduced in \cite{LHCb:2018roe} is
\begin{eqnarray}
\frac{d^4\Gamma}{dq^{2}\;dc_{\theta}\;dc_{\theta_{l}}\;d\phi}=\frac{9}{32\pi}\sum_{i}J_{i}(q^2)f_{i}(\theta, \theta_{l},\phi), \label{eq:17}
\end{eqnarray}
where $c_{\theta}$ represents $\cos{\theta}$. Each of these are mentioned in Table\ref{table 2}. In the given scenario, the angle $\theta$ represents the deviation between the direction of pion emission and the $-\hat{z}$ rest frame Direction of the $D_{s}^{*}$ meson. Similarly, $\theta_{l}$ denotes the angle formed by the $\ell^{-}$ particle with the $+\hat{z}$ rest frame direction of the $\ell^{+}\ell^{-}$ pair. and $\phi$ is the angle made between the planes of the decay as presented in Fig. \ref{fig: Kinematics of quasi four body decay}

The corresponding amplitudes for the particular polarizations have the parametrized form and are the functions of $q^2$ \cite{Schneider}
\begin{eqnarray}
\mathcal{A}^{\perp}_{L,R}\left(q^2\right) &=& -N_{l}\sqrt{2N_{D_{s}^{*}}}\sqrt{\lambda \left(m_{1}^{2},m_{2}^{2},q^2\right)}\left\{\left(C_{9}^{eff}\mp C_{10}\right)\frac{V\left(q^2\right)}{m_{1}+m_{2}}+2\hat{m_{b}}C_{7}^{eff}T_{1}\left(q^2\right)\right\},\notag\\
\mathcal{A}^{\parallel}_{L,R}\left(q^2\right) &=& N_{l}\sqrt{2N_{D_{s}^{*}}}\left\{\left(C_{9}^{eff}\mp C_{10}\right)\left(m_{1}+m_{2}\right)A_{1}\left(q^2\right)+2\hat{m_{b}}C_{7}^{eff}\left(m_{1}^{2}-m_{2}^{2}\right)T_{2}\left(q^2\right)\right\},\notag\\
\mathcal{A}^{0}_{L,R}\left(q^2\right) &=& \frac{N_{l}\sqrt{2N_{D_{s}^{*}}}}{2m_{2}\sqrt{q^2}}\left\{\left(C_{9}^{eff}\mp C_{10}\right)\left[\left(m_{1}^{2}-m_{2}^{2}-q^{2}\right)\left(m_{1}+m_{2}\right)A_{1}\left(q^2\right)-\frac{\lambda\left(m_{1}^{2},m_{2}^{2},q^2\right)}{m_{1}+m_{2}}A_{2}\left(q^2\right)\right]\right.\notag\\
&&\left.+2m_{b}C_{7}^{eff}\left[\left(m_{1}^{2}+3m_{2}^{2}-q^{2}\right)T_{2}\left(q^2\right)-\frac{\lambda\left(m_{1}^{2},m_{2}^{2},q^2\right)}{m_{1}^{2}-m_{2}^{2}}T_{3}\left(q^2\right)\right]\right\},\notag\\
\mathcal{A}^{t}(q^2) &=& 2N_{l}\sqrt{2N_{D_{s}^{*}}}\frac{\sqrt{\lambda\left(m_{1}^{2},m_{2}^{2},q^2\right)}}{\sqrt{q^2}} C_{10}A_{0}\left(q^2\right), \label{eq:18}
\end{eqnarray}
where
\begin{eqnarray}
N_{l} &=& \frac{i\alpha_{e}G_{F}V_{tb}V_{ts}^{*}}{4 \ pi \sqrt2},\notag\\
N_{D_{s}^{*}} &=& \frac{8\sqrt{\lambda}q^{2}}{3 \times 256 \pi^{3}m_{1}^{3}}\sqrt{1-\frac{4m_{l}^2}{q^2}}\mathcal{B}\left(D_{s}^{*} \;\to\; D_{s} \pi\right) \label{eq:19}
\end{eqnarray}
\begin{table}[h]
\centering
\begin{tabular}{|l|l|l|}
\hline\hline
i & $J_{i}\left(q^2\right)$ & $f_{i}\left(\theta, \theta_{l}, \phi\right)$ \\
\hline\hline
1s & $\left(\frac{3}{4}-\hat{m}_{l}^{2}\right)\left(\left|\mathcal{A}_{L}^{\parallel} \right|^2+\left|\mathcal{A}_{L}^{\perp}\right|^2+\left|\mathcal{A}_{R}^{\perp}\right|^2+\left|\mathcal{A}_{R}^{\parallel}\right|^2\right)+4\hat{m}_{l}^2 Re\left[\mathcal{A}_{L}^{\perp}\mathcal{A}_{R^{*}}^{\perp}+\mathcal{A}_{L}^{\parallel}\mathcal{A}_{R^{*}}^{\parallel}\right]$  & $\sin^{2}{\theta}$ \\
\hline
1c & $\left|\mathcal{A}_{L}^{0}\right|^2+\left|\mathcal{A}_{R}^{0}\right|^2+4\hat{m}_{l}^2\left(\left|\mathcal{A}^{t}\right|^2+2 Re\left[\mathcal{A}_{L}^{0}\mathcal{A}_{R^{*}}^{0}\right]\right)$ & $\cos^{2}{\theta}$ \\
\hline
2s & $\beta_{l}^{2}\left(\left|\mathcal{A}_{L}^{\parallel}\right|^{2}-\left|\mathcal{A}_{L}^{\perp}\right|^{2}+\left|\mathcal{A}_{R}^{\parallel}\right|^{2}+\left|\mathcal{A}_{R}^{\perp}\right|^{2}\right)/4$ & $\sin^2{\theta}\cos{2\theta_{l}}$ \\
\hline
2c & $-\beta_{l}^{2}\left(\left|\mathcal{A}_{L}^{0}\right|^2+\left|\mathcal{A}_{R}^{0}\right|^2\right)$ &  $\cos^2{\theta}\cos{2\theta_{l}}$ \\
\hline
3 & $\beta_{l}^{2}\left(\left|\mathcal{A}_{L}^{\perp}\right|^{2}-\left|\mathcal{A}_{L}^{\parallel}\right|^{2}+\left|\mathcal{A}_{R}^{\perp}\right|^{2}-\left|\mathcal{A}_{R}^{\parallel}\right|^{2}\right)/2$ & $\sin^{2}{\theta}\sin^2{\theta_{l}}\cos{2\phi}$ \\
\hline
4 & $\beta_{l}^2 Re\left[\mathcal{A}_{L}^{0}\mathcal{A}_{L^{*}}^{\parallel}+\mathcal{A}_{R}^{0}\mathcal{A}_{R^{*}}^{\parallel}\right]/\sqrt{2}$ & $\sin{2\theta}\sin{2\theta_{l}}\cos{\phi}$ \\
\hline
5 & $\sqrt{2}\beta_{l} Re\left[\mathcal{A}_{L}^{0}\mathcal{A}_{L^{*}}^{\perp}-\mathcal{A}_{R}^{0}\mathcal{A}_{R^{*}}^{\perp}\right]$ & $\sin{2\theta}\sin{\theta_{l}}\cos{\phi}$ \\
\hline
6s & $2\beta_{l} Re\left[\mathcal{A}_{L}^{\parallel}\mathcal{A}_{L^{*}}^{\perp}-\mathcal{A}_{R}^{\parallel}\mathcal{A}_{R^{*}}^{\perp}\right]$ & $\sin^{2}\theta \cos{\theta_{l}}$ \\
\hline
7 & $\sqrt{2}\beta_{l} Im\left[\mathcal{A}_{L}^{0}\mathcal{A}_{L^{*}}^{\parallel}-\mathcal{A}_{R}^{0}\mathcal{A}_{R^{*}}^{\parallel}\right]$ & $\sin{2\theta}\sin{\theta_{l}}\sin{\phi}$ \\
\hline
8 & $\beta_{l}^{2} Im\left[\mathcal{A}_{L}^{0}\mathcal{A}_{L^{*}}^{\perp}-\mathcal{A}_{R}^{0}\mathcal{A}_{R^{*}}^{\perp}\right]/\sqrt{2}$ & $\sin{2\theta}\sin{2\theta_{l}}\sin{\phi}$ \\
\hline
9 & $\beta_{l}^{2} Im\left[\mathcal{A}_{L^{*}}^{\parallel}\mathcal{A}_{L}^{\perp}+\mathcal{A}_{R^{*}}^{\parallel}\mathcal{A}_{R}^{\perp}\right]$ & $\sin^{2}\theta \sin^{2}\theta_{l} \sin{2\phi}$ \\
\hline\hline
\end{tabular}
\caption{The helicity combinations that correspond to different angular distributions.}
\label{table 2}
\end{table}

Regarding the conjugated CP mode, $B_{c}^{+}\to D_{s}^{*+}\left(\to \;D_{s}\pi\right)l^{+}l^{-}$ weak phase conjugations of the CKM  elements lead to
\begin{equation}
\frac{d^4\bar{\Gamma}}{dq^{2}\;dc_{\theta}\;dc_{\theta_{l}}\;d\phi}=\frac{9}{32\pi}\sum_{i}\bar{J}_{i}(q^2)f_{i}\left(\theta, \theta_{l},\phi\right) \label{20}
\end{equation}
with the substitutions as $J_{1\left(c,s\right),2\left(c,s\right),3,4,7} \to \bar{J}_{1\left(c,s\right),2\left(c,s\right),3,4,7}$ and $J_{5,6s,8,9} \to -\bar{J}_{5,6s,8,9}$.
By doing integration over the angles in the domain $\theta \in \left[0,\pi\right], \theta_{l} \in \left[0,\pi\right]$, and $\phi \in \left[0,2\pi\right]$, the differential width becomes only the function of $q^2$ so,
\begin{equation}
    \frac{d\Gamma}{dq^2} = \frac{1}{4}\left(3 J_{1c}+6 J_{1s}-J_{2c}-2J_{2s}\right). \label{eq:21}
\end{equation}
Similar expression results for the conjugated mode of the transition so that the CP-average differential decay width can be written as,
\begin{equation}
    \frac{d\Gamma}{dq^2} = \frac{1}{2}\left(\frac{d\Gamma}{dq^2}+\frac{d\bar{\Gamma}}{dq^2}\right),\label{eq:22}
\end{equation}
To disentangle the CP-conserving and violating effects, CP (angular $S_{i}$ and asymmetry angular $A_{i}$) coefficients  are defined as
\begin{eqnarray}
    S_{i} &=& \frac{J_{i}+\bar{J}_{i}}{d\left(\Gamma+\bar{\Gamma}\right)/dq^2}, \notag\\
    A_{i} &=& \frac{J_{i}-\bar{J}_{i}}{d\left(\Gamma+\bar{\Gamma}\right)/dq^2},\label{eq:23}
\end{eqnarray}
where $J_{i}$'s are the angular coefficients defined in \cite{Li:2023mrj}. Several observables can be constructed from these $J_{i}$ coefficients, and these quasi-four body processes are sensitive to NP. Other physical observables can be calculated such as forward-backward asymmetry $A_{FB}$, longitudinal, transverse polarization fractions $F_{L\left(T\right)}$ of $D_{s}^{*}$ mesons. This we have done in the next few sections.

\section{Observables and form factors}\label{sec:4}
The Standard relation for the CP asymmetry lepton forward-backward asymmetry is,
\begin{equation}
A_{CP}^{FB}\left(q^2\right) =\frac{1}{d\left(\Gamma+\bar{\Gamma}\right)/dq^2}\times \int_{-1}^{1}d\cos{\theta_{l}}\int_{-1}^{1}d\cos{\theta}\int_{0}^{2 \pi}d\phi\frac{d^{4}\left(\Gamma+\bar{\Gamma}\right)}{dq^2 d\cos{\theta}d\cos{\theta_{l}}d\phi} = \frac{3}{4} A_{6}, \label{eq:24}
\end{equation}
while the CP averaged forward-backward asymmetry is,
\begin{eqnarray}
    A_{FB}\left(q^2\right) &= \frac{3}{4} S_{6} \label{eq:25}
\end{eqnarray}
The polarizations are defined as
\begin{eqnarray}
    F_{L} & = \frac{1}{4}\left(3 S_{1c}-S_{2c}\right), \notag\\
    F_{T} & = \frac{1}{2}\left(3 S_{1s}-S_{2s}\right). \label{eq:26}
\end{eqnarray}
We also focus on the Lepton flavour universality (LFU) ratios as these are important to trace out NP, i.e., 
\begin{eqnarray}
 R^{\mu e} &=&\frac{\int_{4\mu^{2}}^{\left(m_{1}-m_{2}\right)^{2}}\frac{d\Gamma\left(B_{c} \to D_{s}^{*}\left(\to D_{s} \pi\right) \mu^{+}\mu^{-}\right)}{dq^2}\, dq^2}{\int_{4\mu^{2}}^{\left(m_{1}-m_{2}\right)^{2}}\frac{d\Gamma\left(B_{c} \to D_{s}^{*}\left(\to D_{s} \pi\right) e^{+}e^{-}\right)}{dq^2}\, dq^2}, \notag\\ \label{eq:27}
 R^{\tau \mu } &=&\frac{\int_{4\mu^{2}}^{\left(m_{1}-m_{2}\right)^{2}}\frac{d\Gamma\left(B_{c} \to D_{s}^{*}\left(\to D_{s} \pi\right) \tau^{+} \tau^{-}\right)}{dq^2}\, dq^2}{\int_{4\mu^{2}}^{\left(m_{1}-m_{2}\right)^{2}}\frac{d\Gamma\left(B_{c} \to D_{s}^{*}\left(\to D_{s} \pi\right) \mu^{+}\mu^{-}\right)}{dq^2}\, dq^2}, \label{eq:28}
\end{eqnarray}
The form factors used in this work are taken from \cite{Li:2023mrj}, in which entities are derived using the covariant light-front quark model, considered best for exploring mesons and baryons. This model requires initial mesons to be on the shell compared to conventional LFQM \cite{LFQM, LFQM1}. They use the $z-$series parametrization form for the form factors to perform analytical continuation\cite{Azzi:2019yne,FCC:2018evy}.
\begin{eqnarray}
           F(q^2) &=& \frac{F\left(0\right)}{1-\frac{q^2}{m_{pole}^2}}\Bigg\{1+a_{1}\left[\left(z(q^2\right)-z\left(0\right)-\frac{1}{3}\left(z\left(q^2\right)^3-z\left(0\right)^3\right)\right]+a_{2}\left[\left(z\left(q^2\right)^2-z\left(0\right)^2\right)+\frac{2}{3}\left(z\left(q^2\right)^3-z\left(0\right)^3\right)\right]\Bigg\},\notag\\ \label{eq:29}
\end{eqnarray}
where $a_{i}$ are the fitting parameters in space-like region $q^2<0$ and $z\left(q^2\right)$ is defined as, 
\begin{equation}
 z\left(q^2\right) = \frac{\sqrt{\left(m_1+m_2\right)^2-q^2}-\sqrt{\left(m_1+m_2\right)^2-\left(m_1-m_2\right)^2}}{\sqrt{\left(m_1+m_2\right)^2-q^2}+\sqrt{\left(m_1+m_2\right)^2-\left(m_1-m_2\right)^2}}.\label{eq:30}
\end{equation}
The values of free parameters $a_{i}$ and all the form factors at $q^2=0$ give the values as in \cite{Li:2023mrj}. Form factors for $B_{c} \;\to\; D_{s}^{*}$ are calculated by using different approaches in \cite{CEPCStudyGroup:2018ghi, Xue, Yue} but here only central values are taken in the present analysis.

\section{New Model Benchmarks and parameters}\label{sec:5}
For the SM Wilson coefficients we use,
$C_{7}^{SM}=-0.305$, $C_{9}^{SM}=4.186$, $C_{10}^{SM}=-4.559$ which are found at $\mu=m_{b}$. By omitting the CKM unitarity, We can determine the CKM elements $V_{ts}V_{tb}$  \cite{Malami}
\begin{equation}
    V_{ts}V_{tb}^{*} = -V_{cb}\left[1-\frac{\lambda^2}{2}\left(1-2\bar{\rho}+2i\bar{\eta}\right)\right]+O\left(\lambda^{6}\right), \label{eq:31}
\end{equation}
so, we took
\begin{equation}
    |V_{tb}V_{ts}^{*}|=\left(41.4\pm 0.5\right)\times 10^{-3}. \label{eq:32}
\end{equation}
The lifetime of $B_{c}$ meson $\tau_{B_c}=0.510$ ps is taken from the Particle Data Group, and the branching fraction $\mathcal{B}\left(D_{s}^{*} \to D_{s} \pi\right)=5\times 10^{-2}$.
Finally, we found the CP averaged differential branching ratios are calculated within the $q^2\left(\text{GeV}^{2}\right)$ bin $\left[1.1,6.0\right]$ and other physical observables in the different $q^2 \left(\text{GeV}^{2}\right)$ bins $\left[1.1,6.0\right]$, $\left[6.0,8.0\right]$, $\left[11,12.5\right]$, and $\left[15,17\right]$ for all the three generations of leptons. The branching fractions of electron and muon are of the order of $10^{-8}$ magnitude while for tau in the bin $\left[15,17\right]$, it can reach up to $10^{-9}$. $R^{\mu e}$ is consistent with the SM prediction that is 1.00 and $R^{\tau \mu}=0.3801$. In the region $\left[1.1, 6.0\right]$ $\text{GeV}^{2}$, the branching fraction for electron and muon is
\begin{eqnarray}
    \mathcal{B}\left(B_{c} \to D_{s}^{*}\left(\to D_{s} \pi\right) e^{+}e^{-}\right)_{\left[1.1,6.0\right]}&=& 0.614\times 10^{-8},\notag\\
    \mathcal{B}\left(B_{c} \to D_{s}^{*}\left(\to D_{s} \pi\right) \mu^{+}\mu^{-}\right)_{\left[1.1,6.0\right]}&=& 0.616\times 10^{-8}, \notag\\
    \mathcal{B}\left(B_{c} \to D_{s}^{*}\left(\to D_{s} \pi\right) \tau^{+}\tau^{-}\right)_{\left[15.0,17.0\right]}&=& 0.087\times 10^{-9} \label{eq:33}
\end{eqnarray}
This quasi-four-fold distribution provides the number of angular observables to check for NP effects. So now we visit the NP models to trace out the new physics. First of all, we present the Wilson coefficients here we used to trace out the NP, and separation between NP effects is based on the following shifts in values of WCs and mainly the LFU NP contribution to $C_{10}^{\mu}$, these are expressed as \cite{Bharucha}
\begin{eqnarray}
    C^{\mu}_{\left(9,10\right)} &=& C^{U}_{\left(9,10\right)}+C^{V}_{\left(9,10\right)}, \notag\\
    C^{\tau}_{\left(9,10\right)} &=& C^{U}_{\left(9,10\right)}, \label{eq:35}
\end{eqnarray}
the Wilson coefficients $C^{U}_{\left(9,10\right)}$ are linked with the $b \;\to \;s \ell^{+} \ell^{-} \left(\ell= e , \tau\right)$ and $C^{V}_{\left(9,10\right)}$ is associated with $b \;\to\; s \mu^{+} \mu^{-}$ \cite{Alguero:2021anc}. Lepton flavor universal new physics is allowed in particularly $C_{9}^{U}$ as it was a common belief to drop out these as the statistical point of view couldn't justify their presence. Still, their inclusion leads to a new paradigm. Vector couplings to muons are encoded in $C_{9}^{\mu}$. LFUV NP affects only muons $C^{V}_{9e,\tau}=0$. Scenario VIII containing universal coefficient coupling $C_{9}^{U}$ together with the muonic part follows the $SU(2)_{L}$ invariance. These all are independent of external hadron states and their momentum-energy relations.
Here we consider the values obtained from the complete data set and the three prominent $1D$ NP scenarios and eight $D>1$ as presented in Table (\ref{table 3}) and (\ref{table 4}).

Using the NP models mentioned in the above table, we assessed the physical observables in different $q^2$ bins and found significant deviations from the Standard Model results. The calculated branching fractions for the quasi-four body process in the range $1.1\;\text{GeV}^{2} \le q^2 \le 6.0\;\text{GeV}^{2}$ for the electron, muon and in the range $15\;\text{GeV}^{2} \le q^2 \le 17\;\text{GeV}^{2}$ for tau are mentioned in the Table (\ref{table 5}) for all the relevant models. The mentioned scenarios exhibit significant deviations from the SM results imprinting the NP effects.

\begin{table}[h]
\centering
\begin{tabular}{|l|l|l|l|}
\hline\hline
Scenarios & WCs  & 1 $\sigma$ range  & $\Delta \chi^{2}$ \\
\hline
SI & $C_{9}^{U}$ & $-1.08\pm 0.18$ & $27.90$ \\
\hline
SII & $C_{9}^{U}=-C_{10}^{U}$ & $-0.50 \pm 0.12$ & $18.85$ \\
\hline
SIII &$C_{9}^{U}=-C_{9^{\prime}}^{U}$ & $-0.88 \pm 0.16$ & $26.92$\\
\hline\hline
\end{tabular}
\caption{The Wilson coefficients in different new scenarios \cite{Alguero:2021anc}.}
\label{table 3}
\end{table}

\begin{table}[h]
\centering
\begin{tabular}{llllllllll}
\hline
Scenarios & WCs & 1$\sigma$ range & pull &  & Scenarios & WCs &1$\sigma$ range & pull \\
\hline\hline
 & $C_{9 \mu}^{V}$ & $\left[-1.31, -0.53\right]$ & 4.5 &  & - &   $C_{9 \mu}^{V}=-C_{10 \mu}^{V}$ & $\left[-0.27, -0.12\right]$ & 3.6 & \\
SV&   $C_{9}^{U}=C_{10}^{U}$           & $\left[-0.13, 0.58\right]$ & - &  & SIX & $C_{10}^{U}$ & $\left[-0.09, 0.27\right]$ & - &\\
& $C_{10 \mu}^{V}$  &  $\left[-0.66, 0.07\right]$ & \\
\hline
 & $C_{9 \mu}^{V}=- C_{10 \mu}^{V}$ & $\left[-0.33, -0.20\right]$ & 4.1 &  &
  & $C_{9 \mu}^{V}$ & $\left[-0.72, -0.41\right]$ & 4.6\\
SVI&  $C_{9}^{U}=C_{10}^{U}$   &   $\left[-0.43, -0.17\right]$ & & & SX  &$C_{10}^{U}$ &$\left[0.05, 0.34\right]$   & \\
\hline
 & $C_{9 \mu}^{V}$ & $\left[-0.43, -0.08\right]$ & 5.5 &  &  & $C_{9 \mu}^{V}$ & $\left[-0.82, -0.51\right]$ &4.6 \\
SVII&     $C^{U}_{9}$      &   $\left[-1.07, -0.58\right] $ & & &SXI
& $C_{10 ^{\prime}}^{U}$& $\left[-0.26, -0.04\right]$ & \\
\hline
 & $C_{9 \mu}^{V}=- C_{10 \mu}^{V}$ &$\left[-0.18, -0.05\right]$ & 5.6 &  &  & $C_{9 \mu}^{V}$ & $\left[-0.96, -0.60\right]$ &5.1\\
SVIII&  $C_{9}^{U}$ &  $\left[-1.15, -0.77\right]$ & & &SXII &$C_{9^{\prime} \mu}^{V}$ &$\left[0.22, 0.63\right]$ &\\
& & & & & &$C_{10}^{U}$ 
&$\left[0.01, 0.38\right]$ & \\
& & & & & &$C_{10^{\prime}}^{U}$ &$\left[-0.08, 0.24\right]$ & \\
\hline\hline
\end{tabular}
\caption{The Wilson coefficients in different new scenarios \cite{Alguero:2021anc}.}
\label{table 4}
\end{table}
\begin{table}[h!]
    \centering
    \begin{tabular}{rrrrrr}
        \hline\hline
        $\mathcal{B}r\left(\text{GeV}^{-2}\right)$ & $q^2\left(\text{GeV}^{2}\right)$ & SM $\left(10^{-8}\right)$ & SII$\left(10^{-10}\right)$ & SV$\left(10^{-10}\right)$ & SVI$\left(10^{-10}\right)$ \\
        \hline
        $\ell=e$ & $\left[1.1, 6.0\right]$ & 0.614 & $\left[5.418,4.532\right]$ & $\left[4.730,4.007\right]$ & $\left[4.064,4.686\right]$ \\
        \hline
        $\ell=\mu$ & $[1.1, 6.0]$ & 0.616 &  & $\left[4.553,8.901\right]$ & $\left[4.267,5.169\right]$\\
        \hline
       $\ell=\tau$ & $\left[15, 17\right]$ & 0.087 & $\left[0.107,0.913\right]$ & $\left[0.646,0.729\right]$ & $\left[0.749,0.925\right]$\\
        \hline\hline
    \end{tabular}
    \caption{Branching ratios in the bin $\left[1.1, 6.0\right]\text{GeV}^{2}$ for SM and different NP models.}
    \label{table 5}
\end{table}
\begin{figure}[!htb]
\centering\scalebox{1}{
\begin{tabular}{ccc}
\includegraphics[width=2.0in,height=1.5in]{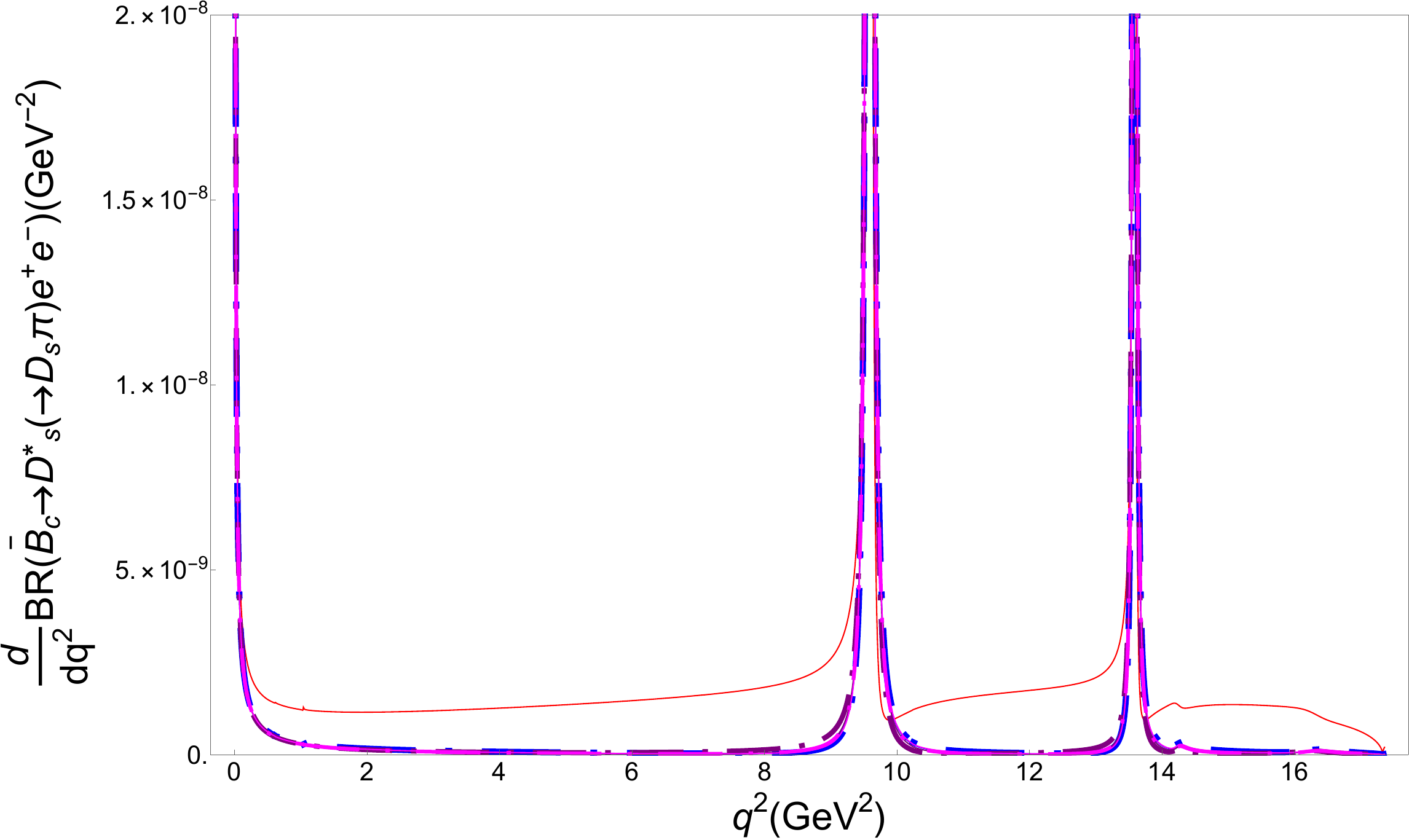}\hspace{0.5cm}\qquad & \qquad
\includegraphics[width=2.0in,height=1.5in]{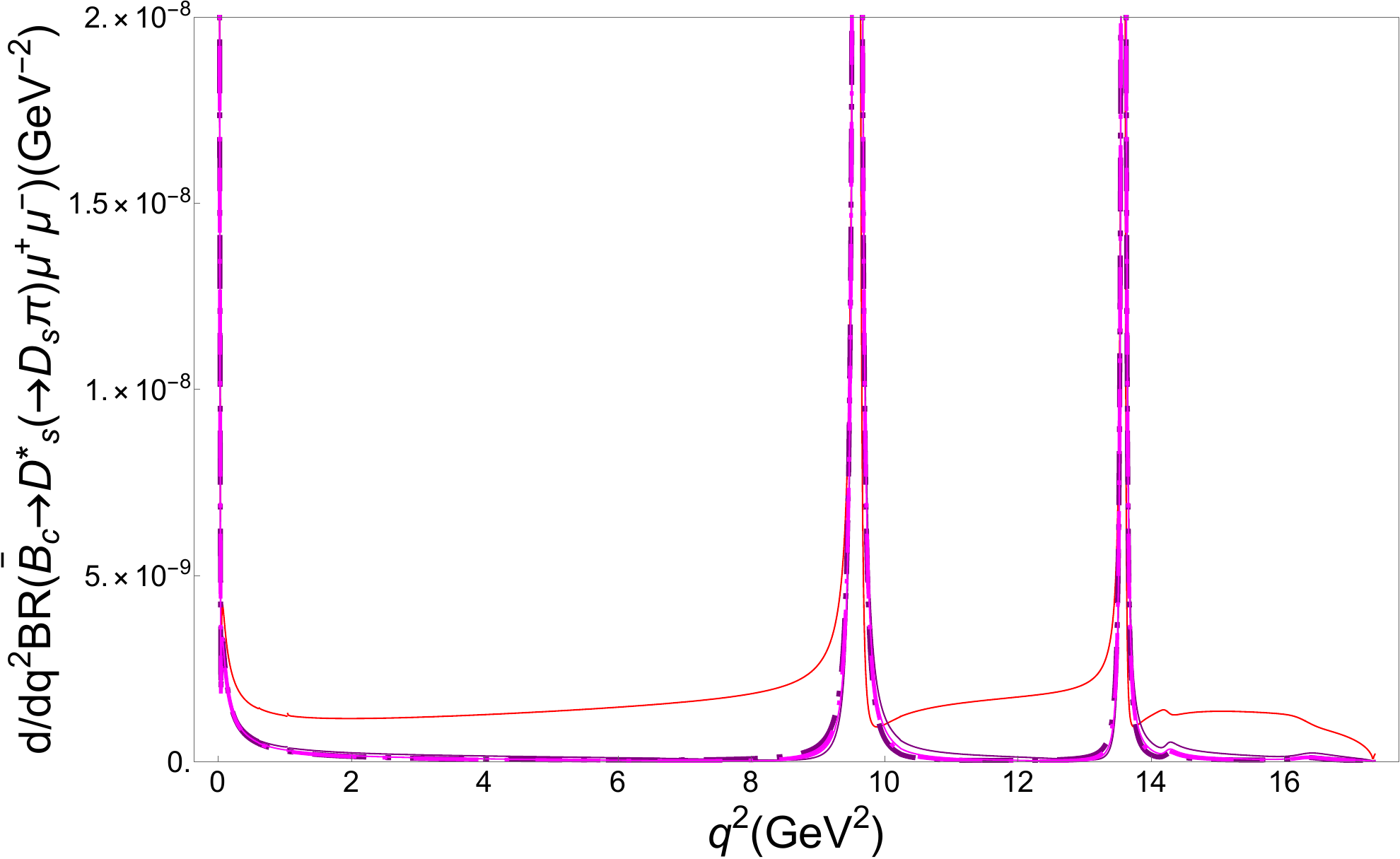}\qquad & \qquad
\includegraphics[width=2.0in,height=1.5in]{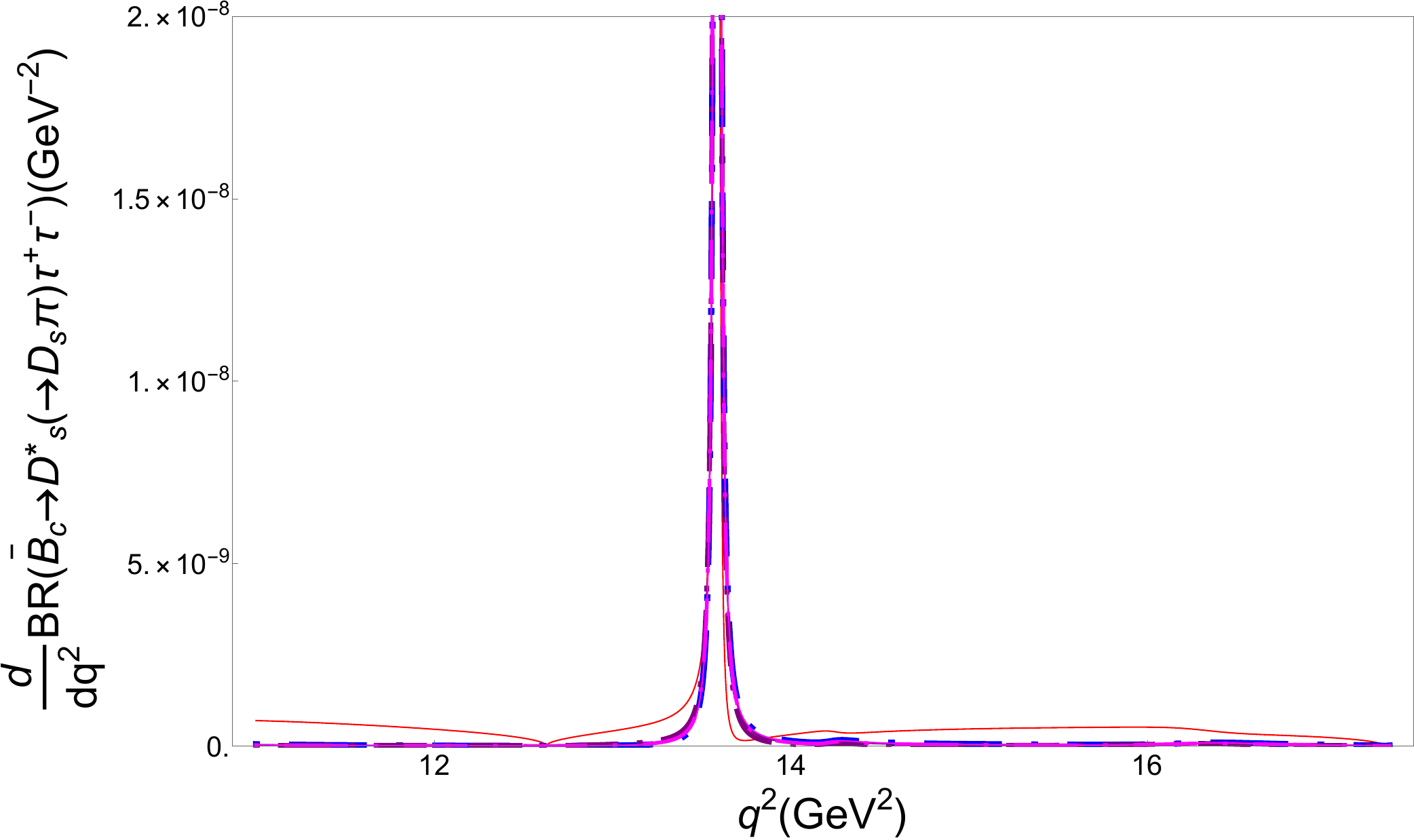} \\
(a)\hspace{0.5cm}\qquad &  \qquad (b)\qquad &  \qquad (c)
\end{tabular}}
\caption{The branching ratios of $B\to D_{s}^*\left(\to D\pi\right)\ell^{+}\ell^{-}$, for $\ell = e,\; \mu,\; \tau$ as a function of $q^2$ in the SM and various NP scenarios is represented where red (solid line) is representing SM, blue (solid line) is representing SII upper interval, blue (dashed) is representing SII lower interval, purple (solid) is representing SV upper interval, purple (dashed) is representing SV lower interval, magenta (solid) is representing SVI upper interval and magenta (dashed) is representing SVI lower interval of the Wilson coefficients. }
\label{Fig1}
\end{figure}

Resonanant particles, charmonium states $J/\psi$ and $\psi(2S)$ affect the widths widely because of large dilepton masses and small widths, while those above the $D\bar{D}$ threshold due to their larger widths don't contribute effectively. Charmless vector mesons are suppressed by the factor $\lambda_{u}$. 

We have investigated the physical observables, the lepton forward-backward asymmetry parameter, transverse and longitudinal polarization fractions. These observables (Average value) are calculated by using the following equation,
\begin{eqnarray}
    \langle A \rangle =\frac{\int_{q^2_{min}}^{q^2_{max}}A\left[q^2\right]\left(\frac{d\Gamma}{dq^2}+\frac{d\bar\Gamma}{dq^2}\right)\,dq^2}{\int_{q^2_{min}}^{q^2_{max}}\left(\frac{d\Gamma}{dq^2}+\frac{d\bar\Gamma}{dq^2}\right)\, dq^2} \label{eq:34}
\end{eqnarray}

 Now we mention the numerical results of these observables, i.e., lepton forward-backward asymmetry and polarization fractions in Tables \ref{table 6} and \ref{table 7} for the respective decay for all the relevant NP models in different $q^2$ bins and the impact of LFUV-NP and LFU NP can be observed as compared to SM results. It's clear from the results that $R_{K}$ and $R_{K^*}$ deviate from the corresponding ratios in SM, signaling the presence of theory beyond the Standard Model. The $q^2$ dependence of these observables is represented in Figures \ref{Fig1} and \ref{Fig4}). The results depict the discrepancy between the results of NP and SM with large deviations showing that $A_{FB}, F_{L}$ and $F_{T}$ are the best probes to search for the physics beyond the SM in different $q^2$ intervals for the specific scenarios mentioned in the Table (\ref{table 4}). Although no such measurements have been recorded experimentally, in the future, it can be expected to reveal the details of this quasi-four body decay by LHCb to measure the corresponding observables. The ratios of branching fractions in different models deviate from 1 in the case of electron and muon, representing that electron and muon are different particles under the constraints of different coupling strengths. Relative deviations of the observables of the respective models with the SM model are calculated. For example, we found a 91.9 \% deviation of the NP branching fraction from the SM value, for $F_{L}$, the relative deviation is 91.3 \%.
 \begin{table}[h]
\centering
\begin{tabular}{lllll}
\hline
$\left[15, 17\right]$ \text{GeV}$^{2}$ & SM & SII & SV & SVI \\
\hline
 $R^{\tau e}$& 0.3801 & $\left[0.0589, 0.6627\right]$ & $\left[0.5790, 0.6748\right]$ & $\left[0.6706, 0.6351\right]$ \\
\hline
\end{tabular}
\caption{Table showing value of  $R^{\tau e}$ in $q^2$ bin of $\left[15, 17\right]$  
 $\text{GeV}^{2}$ in different Scenarios.}
\label{sample_table}
\end{table}

\begin{figure}[!htb]
\centering\scalebox{1}{
\begin{tabular}{ccc}
\includegraphics[width=2.0in,height=1.5in]{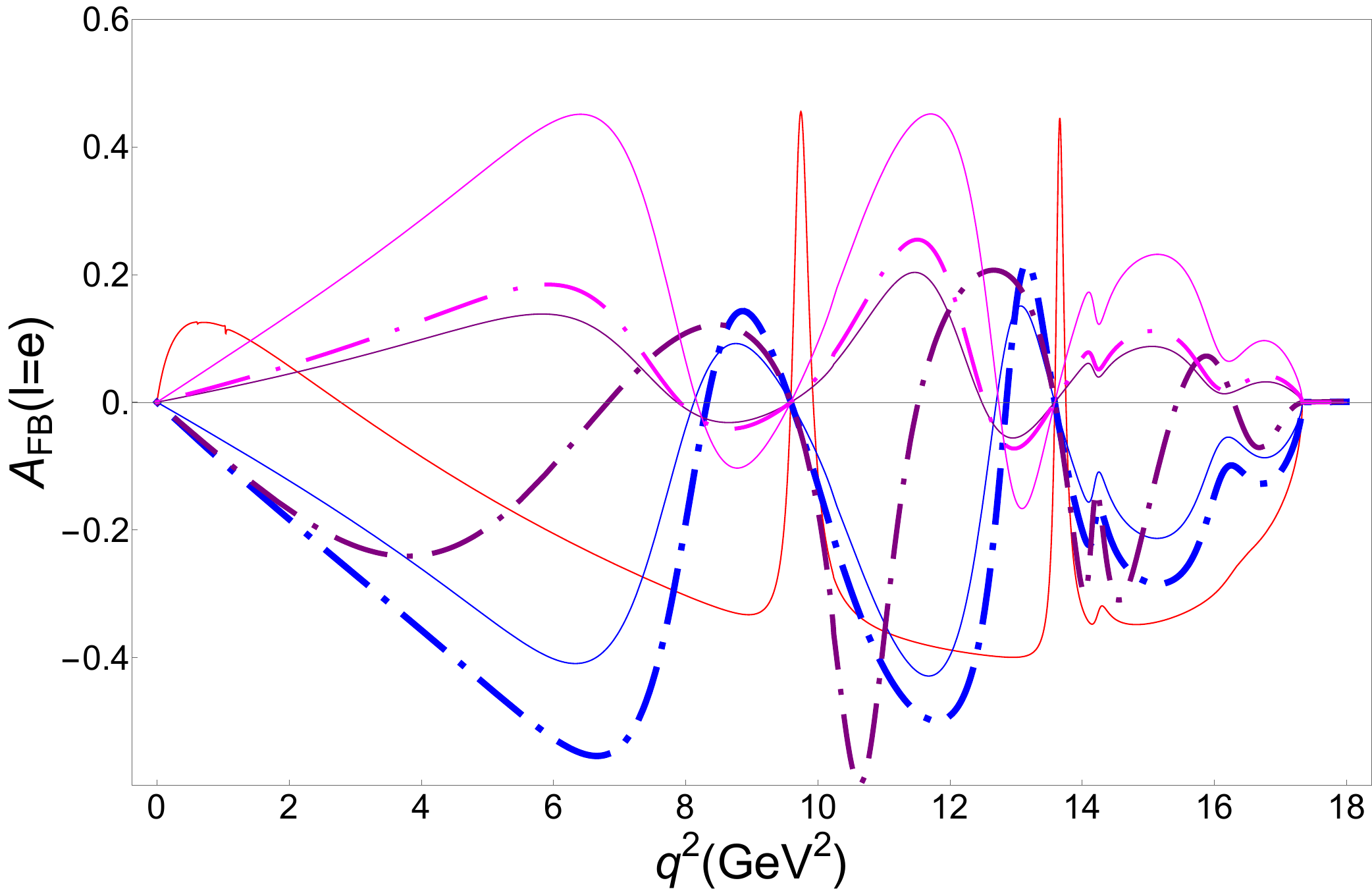}\hspace{0.5cm}\qquad & \qquad
\includegraphics[width=2.0in,height=1.5in]{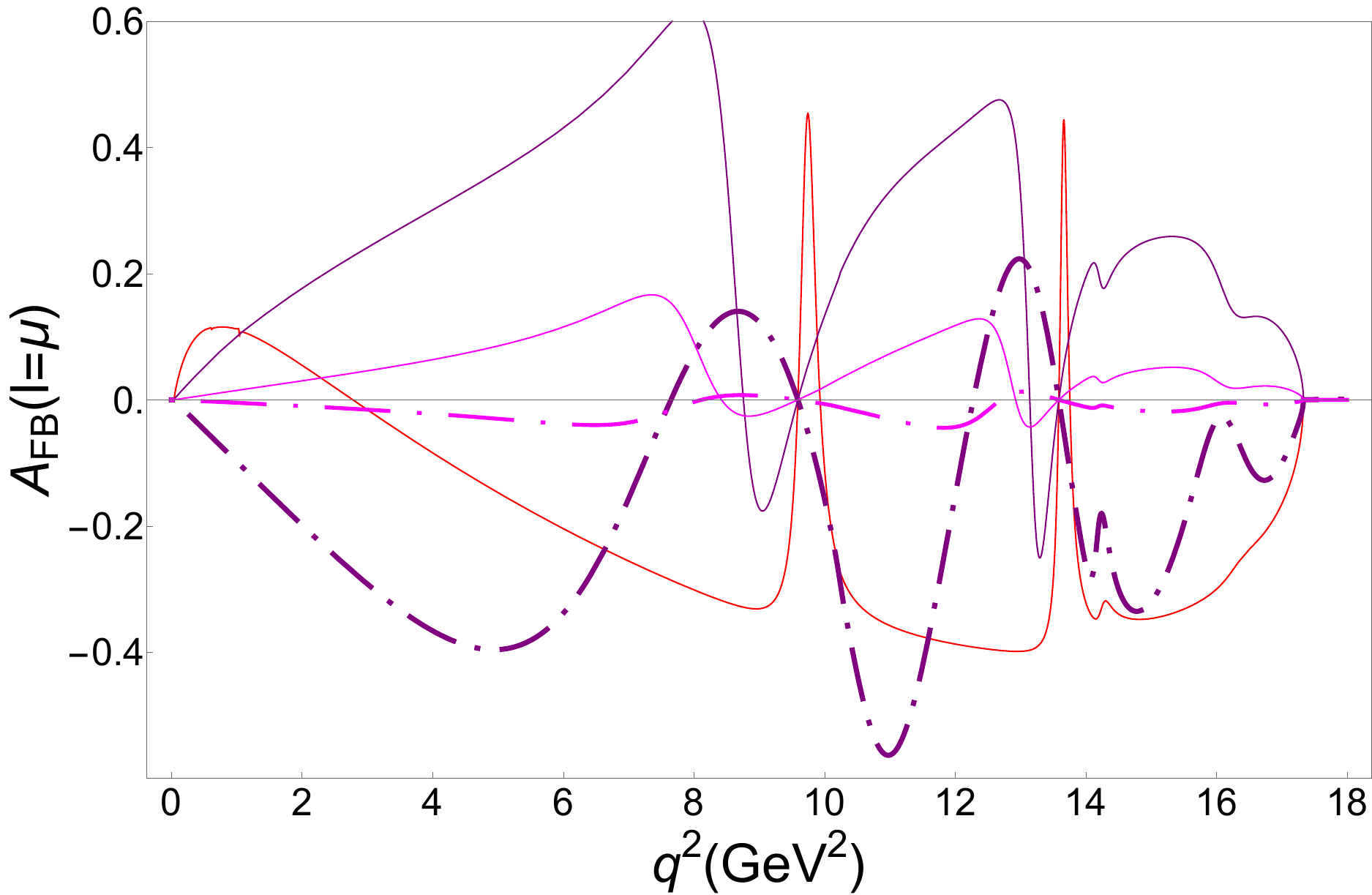}\qquad & \qquad
\includegraphics[width=2.0in,height=1.5in]{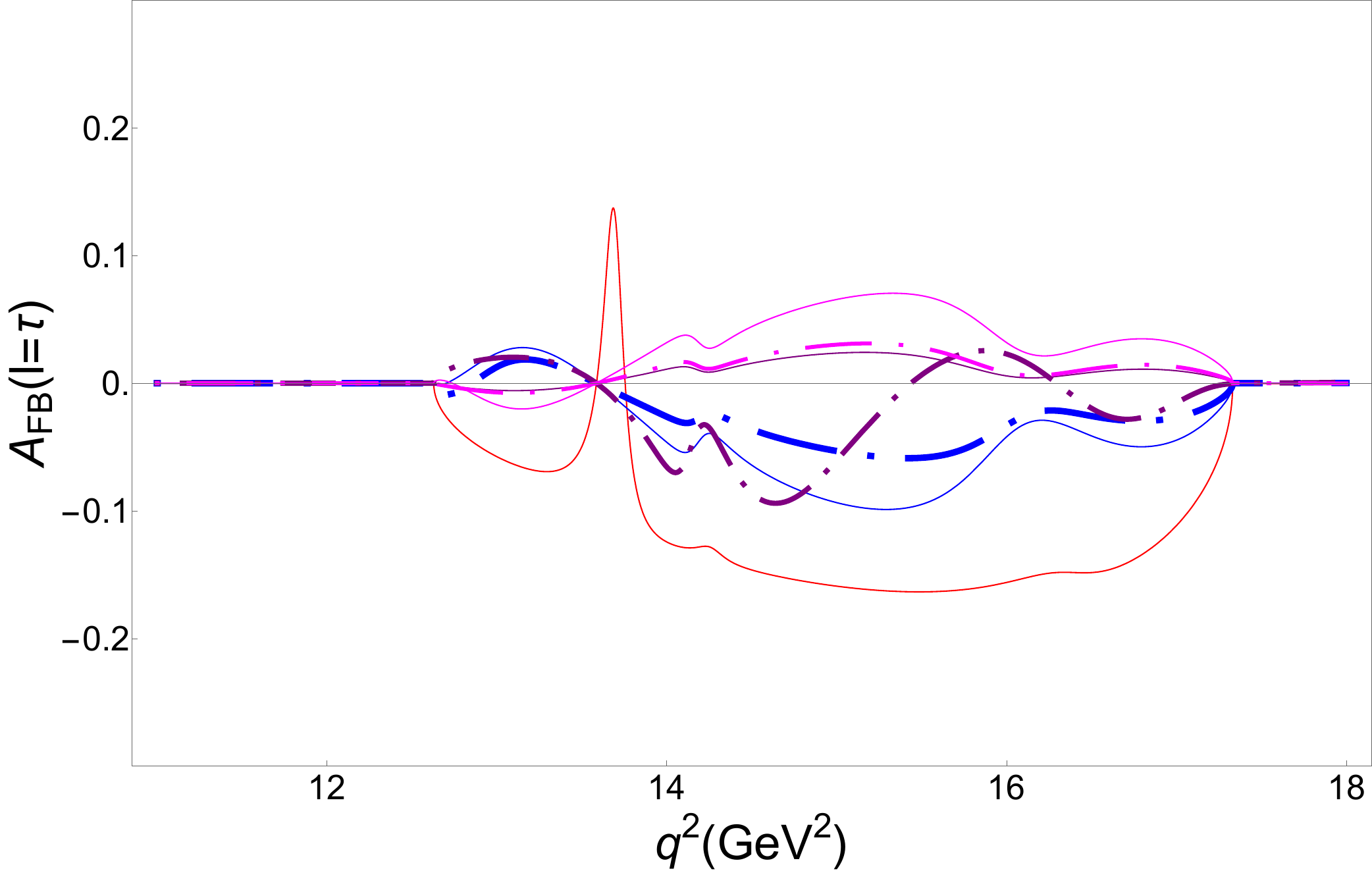} \\
(a)\hspace{0.5cm}\qquad &  \qquad (b)\qquad &  \qquad (c)
\end{tabular}}

\label{Fig2}
\end{figure}
\begin{figure}[!htb]
\centering\scalebox{1}{
\begin{tabular}{ccc}
\includegraphics[width=2.0in,height=1.5in]{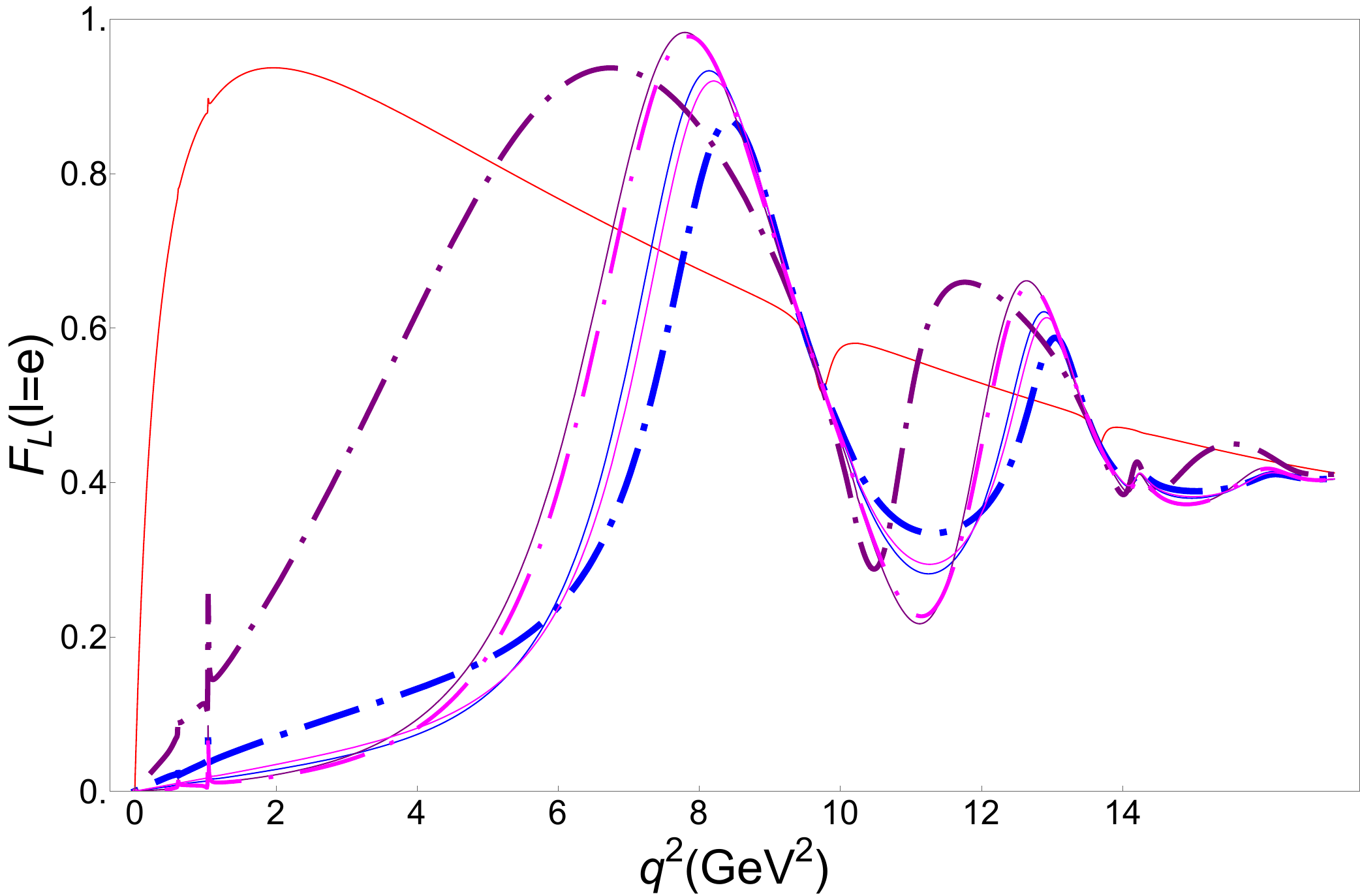}\hspace{0.5cm}\qquad & \qquad
\includegraphics[width=2.0in,height=1.5in]{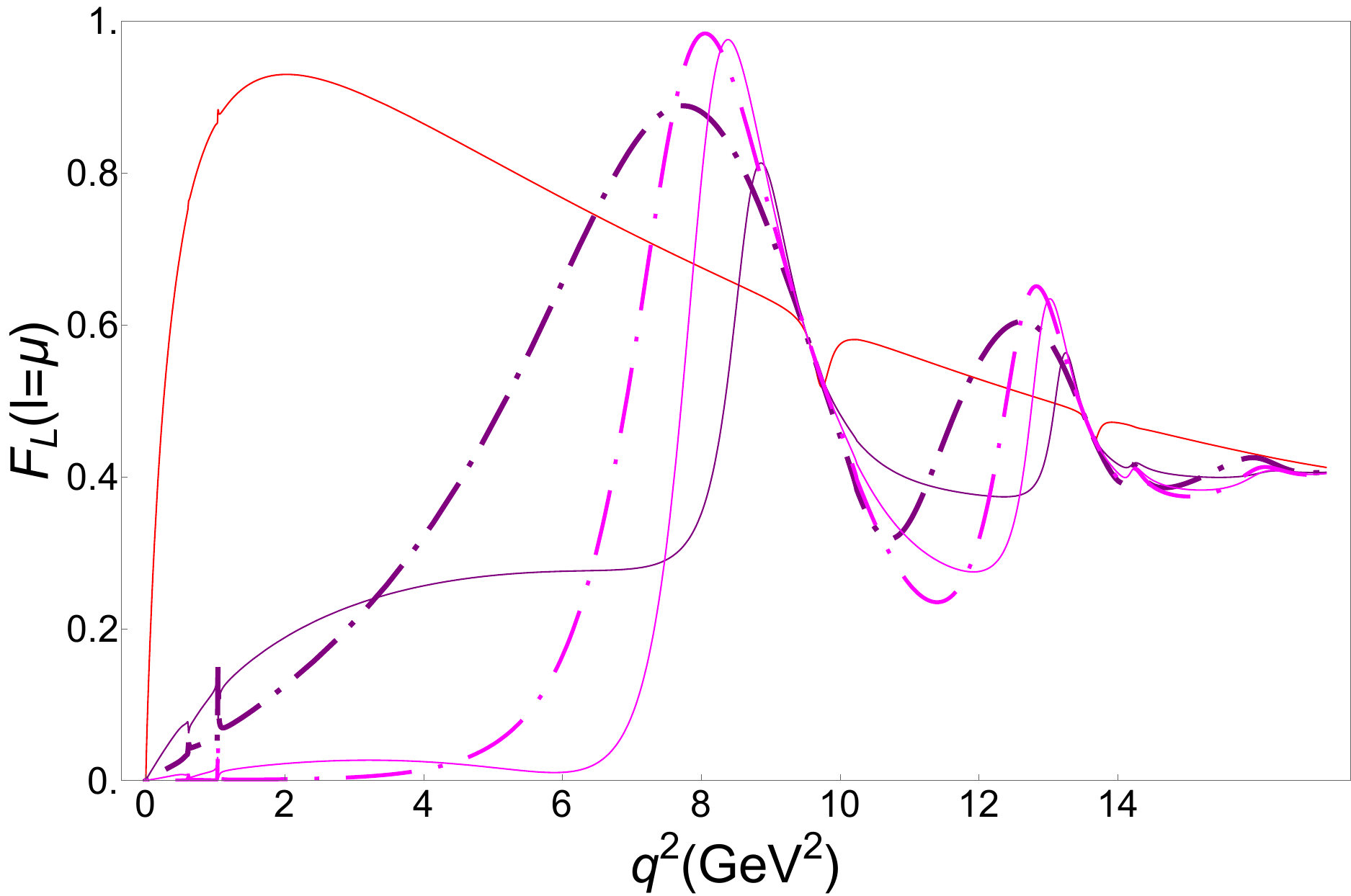}\qquad & \qquad
\includegraphics[width=2.0in,height=1.5in]{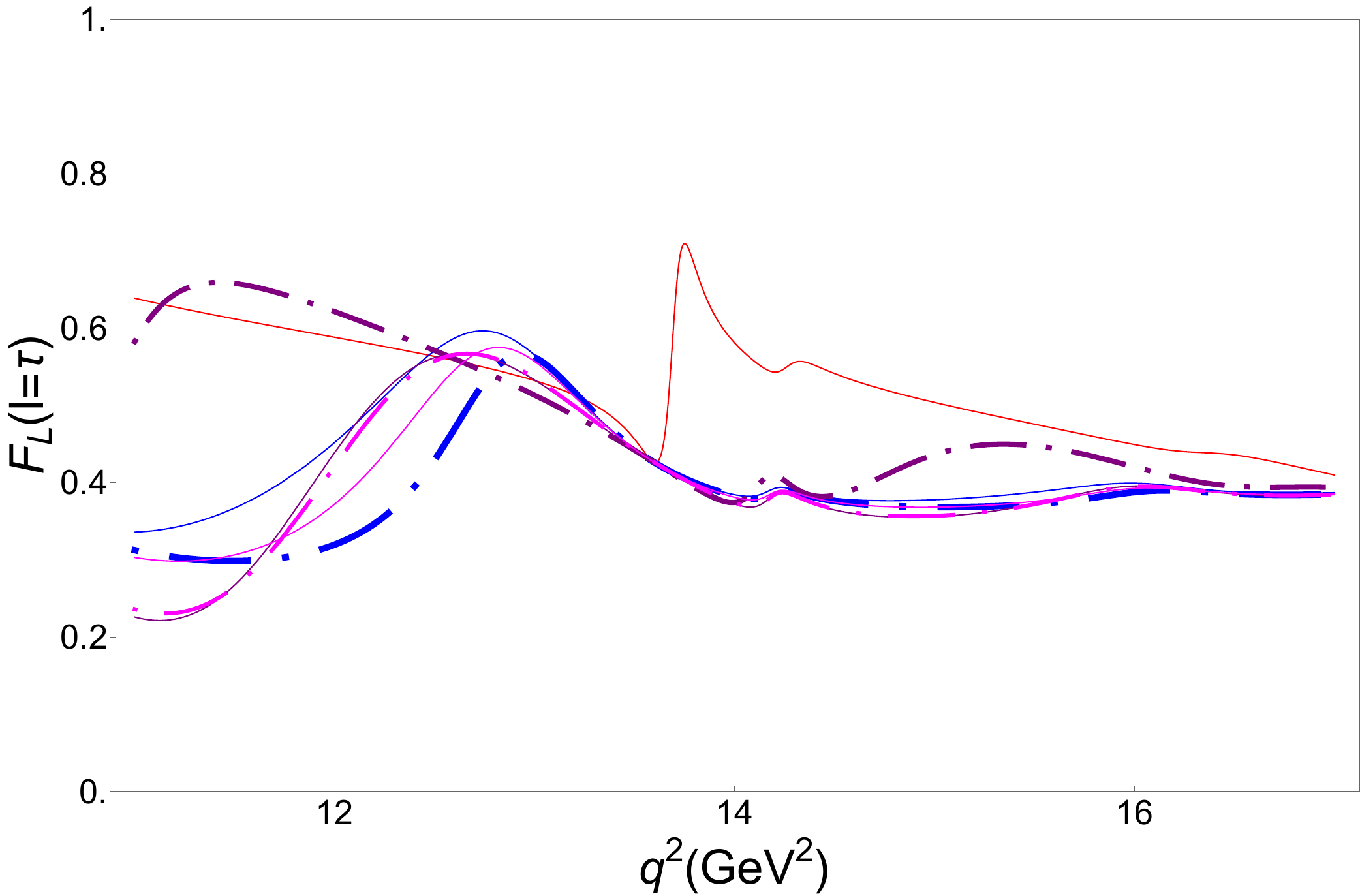} \\
(d)\hspace{0.5cm}\qquad &  \qquad (e)\qquad &  \qquad (f)
\end{tabular}}
\label{Fig3}
\end{figure}
\begin{figure}[!htb]
\centering\scalebox{1}{
\begin{tabular}{ccc}
\includegraphics[width=2.0in,height=1.5in]{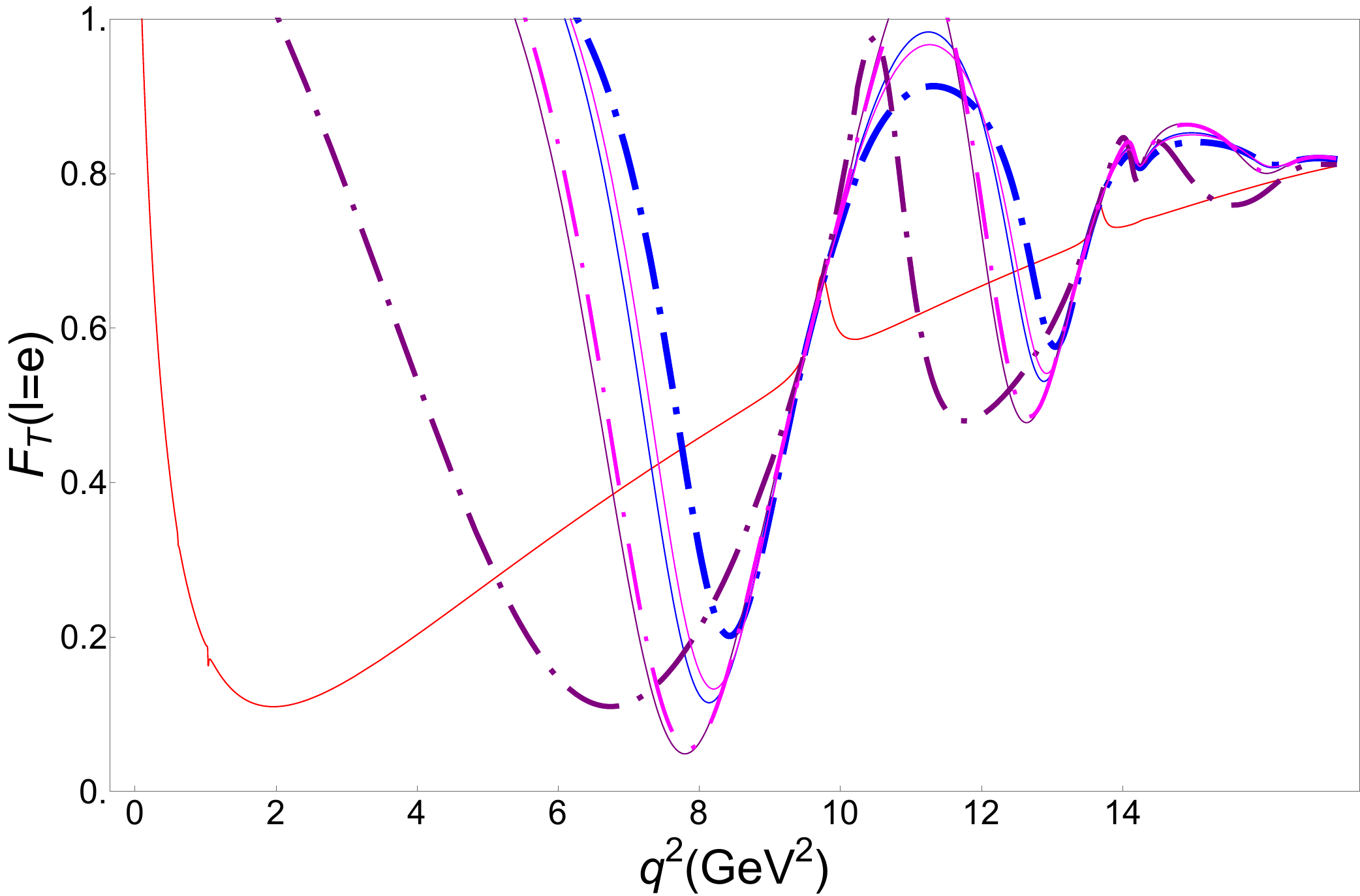}\hspace{0.5cm}\qquad & \qquad
\includegraphics[width=2.0in,height=1.5in]{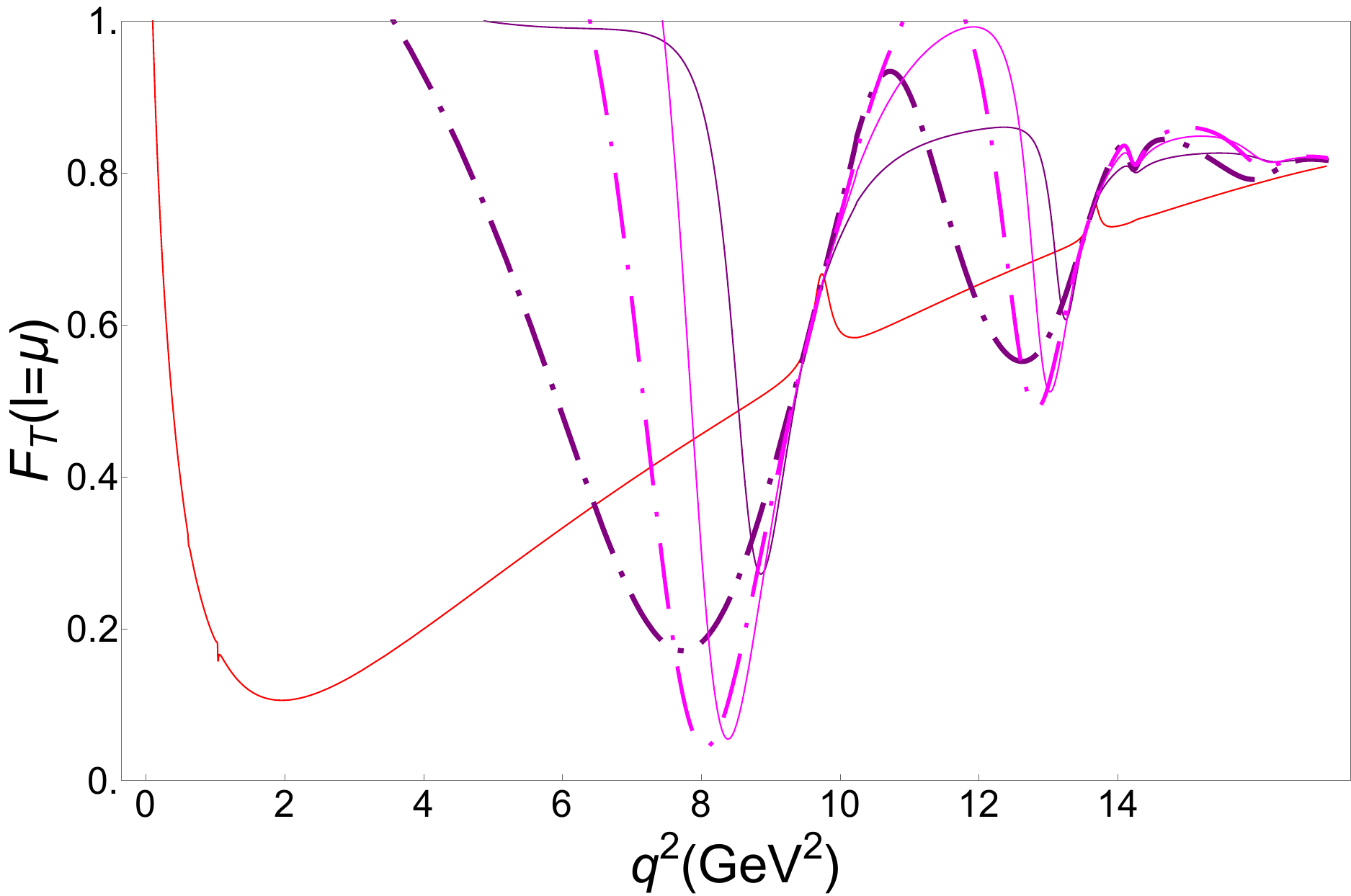}\qquad & \qquad
\includegraphics[width=2.0in,height=1.5in]{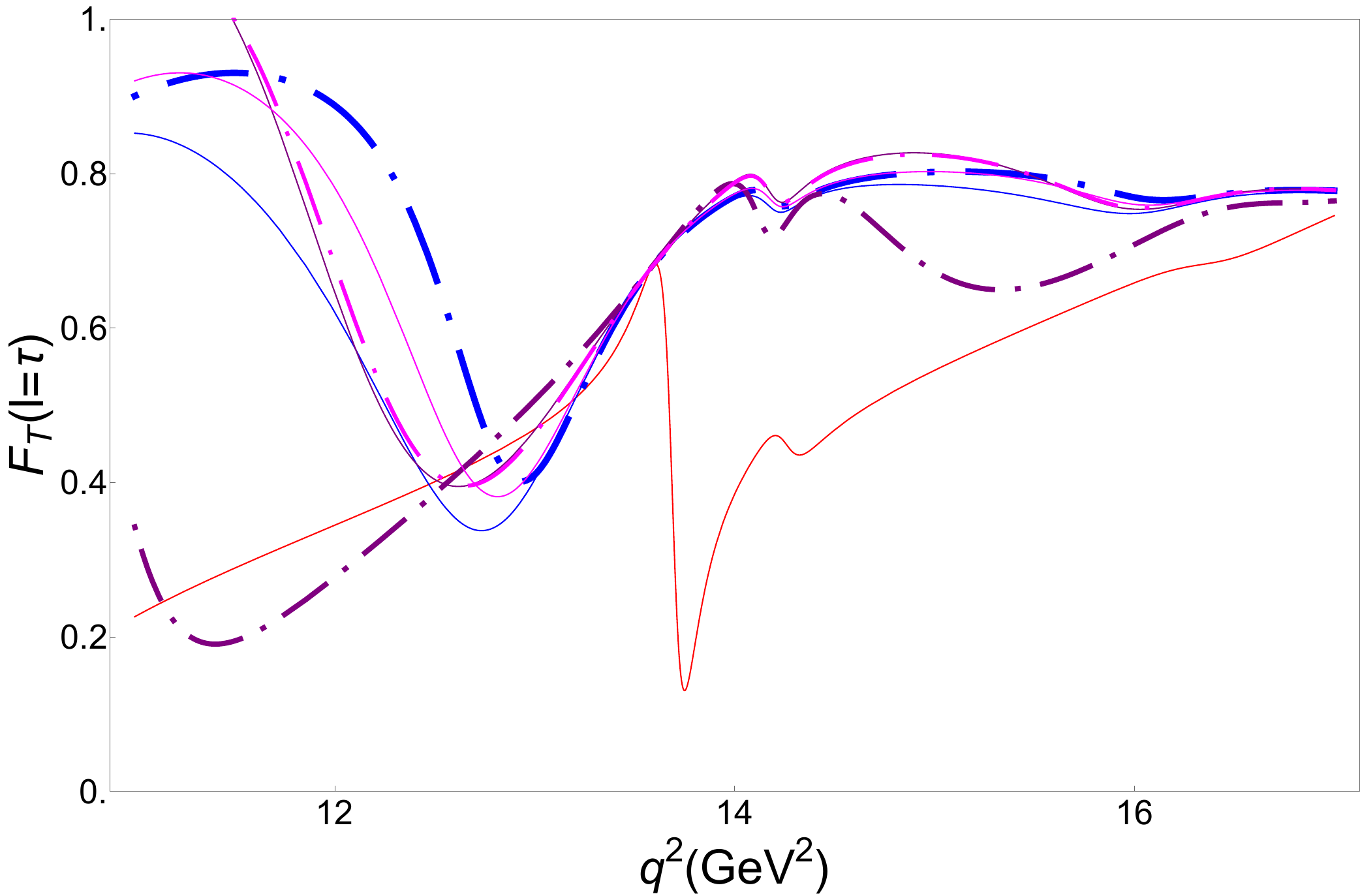} \\
(g)\hspace{0.5cm}\qquad &  \qquad (h)\qquad &  \qquad (i)
\end{tabular}}
\caption{The forward-backward asymmetry (a,b,c), Longitudinal (d, e, f) and transverse polarization (g, h, i) of $B\to D_{s}^*\left(\to D\pi\right)\ell^{+}\ell^{-}$, for $\ell = e,\; \mu,\; \tau$ as a function of $q^2$ in the SM and various NP scenarios is represented where red(solid line) is representing SM, blue (solid line) representing SII upper interval, blue (dashed) representing SII lower interval, purple (solid) representing SV upper interval, purple (dashed) is representing SV lower interval, magenta (solid) is representing SVI upper interval and magenta (dashed) is representing SVI lower interval of the Wilson coefficients. }
\label{Fig4}
\end{figure}
All the other relative deviations concerning Standard Model results can be calculated easily. The CP averaged coefficients $S_{1c(s)}, S_{2c(s)}$ and $S_{6}$ serve as the best probes to search for the BSM. Graphs are depicting variations of the $F_{T}, F_{L}$ and $A_{FB}$ with the $q^2$ and in Table (\ref{table 6}) and (\ref{table 7}), are presented results across different $q^2$ bins.
\begin{table}[h!]
    \centering
    \begin{tabular}{rrrrrr}
        \hline
        Observables & $q^2$ GeV$^2$ & SM & SII & SV & SVI \\
        \hline
        $A_{FB}\left(\ell=e\right)$ & $\left[1.1, 6.0\right]$ & -0.057 & $\left[-0.257,-0.175\right]$ &$\left[-0.178,-0.062\right]$  & $\left[0.081,0.194\right]$ \\
         & $\left[6, 8\right]$ & -0.257 & $\left[-0.484, -0.300\right]$ & $\left[0.029, 0.068\right]$ & $\left[0.098, 0.347\right]$ \\
         & $\left[11, 12.5\right]$ & -0.381 & $\left[-0.463, -0.377\right]$ & $\left[0.078, 0.148\right]$ & $\left[0.194, 0.403\right]$ \\
         & $\left[15, 17\right]$ & -0.294 & $\left[-0.174, -0.113\right]$ & $\left[-0.004, 0.036\right]$ & $\left[0.049, 0.127\right]$ \\
         \hline
         $A_{FB}\left(\ell=\mu\right)$& $\left[1.1, 6.0\right]$& -0.056 & - & $\left[-0.253, 0.244\right]$ & $[-0.014, 0.044]$ \\
         & $[6, 8]$ & -0.256 & - & $\left[-0.125, 0.513\right]$ & $[-0.003, 0.140]$ \\
         & $\left[11, 12.5\right]$ & -0.380 & - &$ \left[-0.263, 0.393\right]$ &  $\left[-0.037, 0.098\right]$ \\
         & $\left[15, 17\right]$ & -0.294 & - & $\left[-0.121, 0.192\right]$ &  $\left[-0.009, 0.032\right]$ \\
         \hline
        $A_{FB}\left(\ell=\tau\right)$ & $\left[1.1, 6.0\right]$ & 0 & 0 & 0 & 0 \\
         & $\left[6, 8\right]$ & 0 & 0 & 0 & 0 \\
         & $\left[11, 12.5\right]$ & 0 & 0 & 0 & 0 \\
         & $\left[15, 17\right]$ & -0.154 & $\left[-0.036, -0.056\right]$ & $\left[-0.002, 0.010\right]$ & $\left[0.014, 0.040\right]$ \\
        \hline
    \end{tabular}
    \caption{Results of $A_{FB}$ in different $q^2$ bins for the three generations of leptons.}
    \label{table 6}
\end{table}

\newpage
\section{Summary}
The rare semi-leptonic decays have shown some deviations from the SM results, motivating us to see if we can find some suitable way to interpret them. Using the latest measurements of $B_s \to \mu^{+}\mu^{-}$, $B_s\to \phi \mu^{+}\mu^{-}, \; R_{K,K_S}$ and $R_{K^*}$ and adding the 254 observables, Alguero \textit{et al.} \cite{Alguero:2021anc} found the pattern of the NP that successfully explain the data. Here, we studied the effect of these new couplings on the angular profiles of the decay $B_{c}\to \left(D_{s}^{*}\to D\pi\right)\ell^{+}\ell^{-}$, where we have studied the four folded decay distribution of this decay in detail and extracted the various possible physical observables from the different angular coefficients. Central values are taken for the form factors, so uncertainties associated with the FFs might show differences in the future if considered. Being exclusive decays, the initial and final state matrix elements are parameterized in terms of the form factors, which are non-perturbative quantities.  We took the values of these form factors calculated in covariant light-front quark model \cite{Li:2023mrj}, where the cascade decays with the $e$ and $\mu$ in the final state have been calculated with the branching ratio to be of the order of  $10^{-8}$. Due to the phase space suppression, for the $\tau$ case, the corresponding result is $\mathcal{O}\left(10^{-9}\right)$ in $q^2\equiv \left[15, 17\right]$\;GeV$^{2}$. When we used the new physics Wilson coefficients from \cite{Alguero:2021anc}, this lepton flavor universality ratio with $\tau$ to $\mu$ showed substantial deviations from the SM results. The case is the same for the other observables, i.e., the leptons forward-backward asymmetry and polarization asymmetries. Hence, these can serve as potential probes to search the New Physics. We hope that in the future, the experimental observations of these rare semi-leptonic decays at the LHCb and dedicated B-factories will help us to find suitable interpretations of these mismatches with the SM results.

\begin{table}[h!]
    \centering
    \begin{tabular}{rrrrrr}
        \hline
        Observables & $q^2\left(\text{GeV}^2\right)$ & SM & SII & SV & SVI \\
        \hline
        $\langle F_{L}\left(\ell=e\right) \rangle$ &$\left [1.1, 6.0\right]$ & 0.870 & $\left[0.099, 0.052\right]$  &
        $\left[0.431, 0.060\right]$ &  $\left[0.054, 0.058\right]$\\
         $\langle F_{T}\left(\ell=e\right) \rangle$& - & 0.130 & $\left[0.901, 0.948\right]$ &$\left[0.569, 0.940\right]$  & $\left[0.946, 0.942\right]$ \\
         \hline
        $\langle F_{L}\left(\ell=e\right)\rangle$ & $\left[6, 8\right]$ & $\left[0.719\right]$ & $\left[0.433, 0.582\right]$ &
        $\left[0.913, 0.805\right]$ & $\left[0.773, 0.540\right]$ \\
        $\langle F_{T}\left(\ell=e\right) \rangle$ & - & 0.281 & $\left[0.567, 0.418\right]$ & $\left[0.087, 0.195\right]$ & $\left[0.227, 0.460\right]$  \\
        \hline
        $\langle F_{L}\left(\ell=e\right)\rangle$ & $\left[11, 12.5\right]$ & 0.535 & $\left[0.353, 0.331\right]$ & $\left[0.635, 0.369\right]$ & $\left[0.354, 0.334\right]$ \\
        $\langle F_{T}\left(\ell=e\right)\rangle$ & - & 0.465 & $\left[0.647, 0.669\right]$ & $\left[0.365, 0.631\right]$ & $\left[0.646, 0.666\right]$  \\
        \hline
        $\langle F_{L}\left(\ell=e\right)\rangle$ & $\left[15, 17\right]$ & 0.431 & $\left[0.401, 0.400\right]$ & 
       $ \left[0.430, 0.404\right]$ & $\left[0.403, 0.401\right]$ \\
        $\langle F_{T}\left(\ell=e\right)\rangle$ & - & 0.569 & $\left[0.599, 0.600\right]$ & $\left[0.570, 0.596\right]$ & $\left[0.597, 0.599\right]$ \\
        \hline
        $\langle F_{L}\left(\ell=\mu\right)\rangle$ & $\left[1.1, 6.0\right]$ & 0.867 & - & $\left[0.219, 0.218\right]$ & $\left[0.012, 0.021\right]$ \\
        $\langle F_{T}\left(\ell=\mu\right)\rangle$ & - & 0.133 & - &-  & - \\
        \hline
        $\langle F_{L}\left(\ell=\mu\right)\rangle$ & $\left[6, 8\right]$ & 0.719 & - & $\left[0.822, 0.284\right]$ & 
        $\left[0.574, 0.146\right]$ \\
        $\langle F_{T}\left(\ell=\mu\right)\rangle$ & - & 0.281 &-  &  -&  -\\
        \hline
        $\langle F_{L}\left(\ell=\mu\right)\rangle$ & $\left[11, 12.5\right]$ & 0.536 &  & $\left[0.494, 0.383\right]$ &  $\left[0.286, 0.292\right]$ \\
        $\langle F_{T}\left(\ell=\mu\right)\rangle$ & - & 0.464 & - & - & - \\
        \hline
        $\langle F_{L}\left(\ell=\mu\right)\rangle$ & $\left[15, 17\right]$ & 0.432 & - & $\left[0.412, 0.403\right]$&  $\left[0.399, 0.398\right]$ \\
        $\langle F_{T}\left(\ell=\mu\right)\rangle$ & - & 0.568 & - & - & - \\
        \hline
        $\langle F_{L}\left(\ell=\tau\right)\rangle$ &$ \left[1.1, 6\right]$ & - &-  & - & - \\
        $\langle F_{T}\left(\ell=\tau\right)\rangle$ & - & - & - & - & - \\
        \hline
        $\langle F_{L}\left(\ell=\tau\right)\rangle$ & $\left[6, 8\right]$ & - & - & - & - \\
        $\langle F_{T}\left(\ell=\tau\right)\rangle $ & - & - &-  & - &  -\\
        \hline
        $\langle F_{L}\left(\ell=\tau\right)\rangle$ & $\left[11, 12.5\right]$ & - & - & - & - \\
        $\langle F_{T}\left(\ell=\tau\right)\rangle$ & - & - &- & - & - \\
        \hline
        $\langle F_{L}\left(\ell=\tau\right)\rangle$ & $\left[15, 17\right] $& 0.458 & $\left[0.381, 0.389\right]$ & 
       $ \left[0.414, 0.383\right]$ & $\left[0.382, 0.383\right]$ \\
        $\langle F_{T}\left(\ell=\tau\right)\rangle$ & - & 0.542 & $\left[0.619, 0.611\right]$ &$ \left[0.586, 0.617\right]$ & $\left[0.618, 0.617\right]$ \\
        
        \hline
    \end{tabular}
    \caption{Results of $F_{L}$ and $F_{T}$ in different $q^2$ bins for the three generations of leptons.}
    \label{table 7}
\end{table}
\newpage
\newpage


\vspace{2cm}
\end{document}